\documentclass[preprint]{elsarticle}
\usepackage{graphicx}
\usepackage{amsmath,amssymb,verbatim,amsthm}
\numberwithin{equation}{section}

\newcommand{\bbE}{\mathbb{E}}

\newcommand{\bbR}{\mathbb{R}}

\newcommand{\bbV}{\mathbb{V}}
\newcommand{\TT}{\mathbb{T}}

\newcommand{\cN}{\mathcal{N}}

\newcommand{\cQ}{\mathcal{Q}}
\newcommand{\cP}{\mathcal{P}}
\newcommand{\cK}{\mathcal{K}}

\newcommand{\rd}{\mathrm{d}}
\newcommand{\Z}{\mathbb{Z}}
\newcommand{\abs}[1]{| #1 |}
\newcommand{\norm}[1]{\| #1 \|}

\newcommand{\R} {\mathbb{R}}

\begin{document}

\title{Proposals which speed-up function-space MCMC}

% \author{
% K.J.H. Law 
% \thanks{
%   Department of Mathematics, University of Warwick, Coventry CV4 7AL, England
% }
% }

%\affiliation{Warwick Mathematics Institute, University of Warwick, Coventry CV4 7AL, UK}}
\author{K.\ J.\ H.\ Law}
% \affil{1}}
%\address{\affilnum{1}
\address{
Warwick Mathematics Institute, University of Warwick,
  Coventry CV4 7AL, UK}

%\address{$^1$ Mathematics Faculty, Open University, Milton Keynes MK7~6AA, UK} \address{$^2$ Department% of Mathematics, Imperial College, Prince Consort Road, London SW7~2BZ, UK} \address{$^3$ Department of C%omputer Science,
%University College London, Gower Street, London WC1E~6BT, UK} 
\ead{k.j.h.law@warwick.ac.uk}

%\begin{abstract}
%{blahby}
%\end{abstract}

%=======================================================================

%=======================================================================

\begin{abstract}
Inverse problems lend themselves naturally to a Bayesian formulation,
in which the quantity of interest is a posterior distribution of state
and/or parameters given some uncertain observations.  For the common
case in which the forward operator is smoothing, then the inverse
problem is ill-posed.  Well-posedness is imposed via regularisation
in the form of a prior, which is often Gaussian.  
Under quite general conditions, it can be shown that the posterior is
absolutely continuous with respect to the prior and it may be
well-defined on function space in terms of its density with respect to 
the prior.  In this case, by constructing a proposal for which the prior is
invariant, one can define Metropolis-Hastings schemes for MCMC which
are well-defined on function space (\cite{Stuart10,CRSW12}), 
and hence do not degenerate as the
dimension of the underlying quantity of interest increases to infinity, e.g. 
under mesh refinement when approximating PDE in finite-dimensions.
However, in practice, despite the attractive theoretical properties of the
currently available schemes, they may still suffer from long
correlation times, particularly if the data is 
very informative about some of the unknown parameters.
In fact, in this case it may be the directions of the posterior which
coincide with the (already known) prior which 
decorrelate the slowest.
The information incorporated into the posterior through
the data is often contained within some finite-dimensional subspace,
in an appropriate basis, perhaps even one defined by eigenfunctions of 
the prior.  We aim to exploit this fact and improve the mixing time of
function-space MCMC by careful rescaling of the proposal.  To this
end, we introduce two new basic methods of increasing
complexity, involving (i) characteristic function truncation of
high frequencies and (ii) Hessian information to interpolate between
low and high frequencies.  The second, more 
sophisticated version, bears some similarities with recent methods
which exploit local curvature information, for example RMHMC 
\cite{girolami2011riemann}, and stochastic Newton \cite{martin2012stochastic}.
These ideas are illustrated with numerical experiments 
on the Bayesian inversion of the heat equation and Navier-Stokes
equation, given noisy observations.

\end{abstract}

\maketitle

\section{Introduction}

Sampling distributions over high-dimensional 
state-spaces is a notoriously difficult problem,
and a lot of effort has been devoted to developing
methods to do this effectively 
\cite{BRSV08, GGR97, robtwe96, CRSW12}.
If a given target distribution is known
pointwise, then
Markov chain Monte Carlo is a general
method which prescribes a Markov chain
having this target distribution as its
invariant distribution. %developed by \cite{}.  
Examples include Metropolis-Hastings
schemes, developed first by \cite{MRRT53} and 
later generalized for arbitrary proposals
by \cite{Has70}.
It is well known that Metropolis-Hastings
schemes developed for finite-dimensional
systems degenerate in the limit of inference
on infinite-dimensional spaces and a lot of 
work has been devoted to examining the actual
rate with which they degenerate \cite{GGR97}.
Indeed this analysis has led to improved performance
of chains in finite dimensions through arguments about
optimal scaling of the acceptance probability.
Recently, it has been shown in a series of 
works by \cite{BRSV08,CRSW12,Stuart10}
that it is possible to overcome this 
limitation by defining the Metropolis-Hastings
scheme in function space, thus confronting
the issue of high dimensions head on.
This is particularly useful, for example, in
the case where the system is a finite-dimensional
approximation of a PDE.  In such a case, 
the convergence properties of the function-space scheme 
are independent of
mesh refinement which increases the dimension of 
the finite-dimensional approximating system.
However, it may (and often does in practice)
happen that, even though the scheme is defined
on function-space and is hence independent of
mesh refinement, the integrated autocorrelation
may be quite large for certain degrees of freedom,
in particular those which have small effect on the 
likelihood.  Hence, there is clearly a paradox in 
that the degrees of freedom which are
not informed by the data, and hence are governed by the prior,
are the rate limiting factor in the convergence of MCMC.
But, at the same time, more is known {\it a priori}
about such degrees of freedom.  We aim to maintain
the positive aspect of being defined in function space
while removing the deficit of mixing poorly in 
directions which are better known a priori.  This will be 
accomplished by 
appropriate rescaling based on curvature which 
leverages a priori known information.
Inspired by the works \cite{BRSV08,CRSW12,Stuart10}
we will focus on sampling distributions $\mu$ over some 
Hilbert space $H$
which have density with respect to a Gaussian
reference measure $\mu_0$ of the following form
\begin{equation}
\mu (du) = Z \exp [-\Phi(u)] \mu_0(du),
\label{density}
\end{equation}
where $\Phi: H \rightarrow \bbR$, and
$Z = 1/ \int_H \exp [-\Phi(u)] \mu_0(du)$.
For simplicity, we let
\begin{equation}
\Phi(u) = \frac{1}{2 \gamma^2} |G(u) - y|^2,
\label{loglike}
\end{equation}
and we will further assume that $G(u)$ 
is a compact operator.
This function arises as the negative log likelihood of
$u \sim \mu_0$ given a realization of an
observation $y = G(u) + \eta$, where 
$\eta \sim \cN(0,\gamma^2I)$ independent of $u$.
The posterior distribution on $u$, conditioned on the 
particular observation $y$, is given by \eqref{density}.
Furthermore, %without any loss of generality, 
we will assume
that $\mu_0 = N(m,C)$ for some trace-class operator $C$.
%The joint distribution of 
%the random variables $(u,y)$ given that 
%$y$ takes this specific value is hen, 
%$\mu$ is t, so $\mu$ is the posterior
%on $u$ conditioned on the given realization of $y$.

The rest of this paper will be organized as follows.  In 
Section \ref{back}, we will provide a review of 
necessary concepts, including the recently developed 
function-space MCMC method pCN \cite{CRSW12} which
will serve as a starting point for this work.  In Section
\ref{proposal} we illustrate the remaining decorrelation 
issues pertaining to proposal chains of the pCN form. 
In Section \ref{operator} we introduce a
class of constant operator rescaled proposals 
which address the issues highlighted in Section 
\ref{proposal} and yet still fit into the function-space
sampler framework, hence taking full advantage of its benefits. 
In Section \ref{numerics}, we investigate the performance 
of these algorithms computationally against standard pCN
for the inverse heat equation and inversion of Navier-Stokes equation.
%in terms of PSRF \cite{brooks1998general} and 
%autocorrelation diagnostics, as well as variance convergence.
Finally, we provide conclusions and possible 
future directions in Section \ref{conclusion}.

\section{Background}
\label{back}

In this section we give some background on Markov chain Monte Carlo (MCMC)
methods, and in particular the Metropolis-Hastings (MH) variants.  
We then 
introduce the preconditioned Crank-Nicolson (pCN) proposal that
yields an MH algorithm which is well-defined on function spaces for a
wide range of posterior distributions of the form \eqref{density}, for
example those arising from the Bayesian interpretation of the solution to
an inverse problem with Gaussian prior and observational noise.
Finally, we will describe the way that correlations enter into
approximations based on the resulting set of samples and define the 
effective sample size, which will be relevant in the derivation of the 
new methods which follow.

\subsection{MCMC}
\label{mcmc}

MCMC methods aim to sample from a target distribution $\mu$ over
$H$ by designing a Markov transition kernel $\cP$ such that 
$\mu$ is invariant under the action of $\cP$
\begin{equation}
\int_H \mu(du) \cP(u,dv) = \mu(dv),
\label{invariant}
\end{equation}
with shorthand ($\mu \cP = \mu$), where the integral on the
lefthand side is with respect to $du$.
The condition known as {\it detailed balance} between a transition
kernel $\cP$ and a probability distribution $\mu$ says that
\begin{equation}
\mu(du) \cP(u,dv) = \mu(dv) \cP(v,du), \quad \forall ~ u,v \in H.
\label{db}
\end{equation}
Integrating with respect to $u$, one can see that 
detailed balance implies $\mu \cP = \mu$.
Metropolis-Hastings methods prescribe an accept-reject move based
on proposals from an arbitrary transition kernel $\cQ$ in order 
to define a kernel $\cP$ such that detailed balance with an arbitrary 
probability distribution  
is guaranteed.  In order to make sense of MH in infinite dimensions,
first define the measures \cite{Tie}
\begin{eqnarray}
\begin{array}{lll}
\nu(du,dv) &=& \cQ(u,dv) \mu(du) \\
\nu^T(du,dv) &=& \cQ(v,du) \mu(dv).
\end{array}
\end{eqnarray}
Provided $\nu^T \ll \nu$, where $\ll$ is 
denoting absolutely continuity when comparing measures, 
one can define the MH kernel as follows.
Given current state $u_n$,
a proposal is drawn $u^* \sim \cQ(u_n,\cdot)$, 
and then accepted with probability
\begin{equation} 
\alpha(u_n,u^*) = {\rm min} 
\left \{1, 
\frac{d\nu^T}{d\nu}(u_n,u^*)
 \right \}.
\label{acceptance} 
\end{equation}
The resulting chain has transition kernel $\cP$
given by %.  Properly speaking
\begin{equation} 
\cP(u,dv) = \alpha(u,v) \cQ(u, dv) + \delta_u(dv) \int_X  (1- \alpha(u,w)) \cQ(u, dw).
\label{mh_transition} 
\end{equation}
So, we have that 
\begin{eqnarray} 
\nonumber
\mu(du) \cP(u,dv) = {\rm min} 
\left \{\cQ(u, dv) \mu(du), \cQ(v,du) \mu(dv) \right \} + \\
{\rm max} 
\left \{0, \delta_u(dv) \left[ \mu(du) - \int_X \cQ(w, du) \mu(dw)
  \right ] \right \}
\end{eqnarray} 
which one can see
satisfies the necessary symmetry condition \eqref{db} .
Under an additional assumption of geometric ergodicity, 
one has that $|\cP^n(u_0,\cdot)-\mu|_{\rm TV} \rightarrow 0$ for any 
$u_0 \in H$ \cite{HSV12}.
Therefore, after a sufficiently long equilibration period, or {\it burn-in},
one has a prescription for generating samples approximately 
distributed according to $\mu$, 
from which integrals with respect to $\mu$
can be approximated.  These samples are correlated, however, 
and this will be revisited below.

Since there is great freedom in choice of proposal kernel $\cQ$,
one can imagine that there is great discrepancy in the performance 
of the algorithms resulting from different choices.  
It is popular to choose symmetric kernels, such as the classical
random walk algorithm, in which the difference between the 
current and proposed states is taken to be a centered Gaussian distribution, 
e.g.
$$
\cQ(u_n,\cdot) = \cN(u_n, \beta^2 C),
$$  
where $C$ is the covariance of the prior $\mu_0$.
%This way the $\cQ$ terms drop out of the expression for the 
%acceptance probability, which may be an attractive simplification.
%But, f
For this proposal, the absolute continuity condition $\nu^T \ll
\nu$ is violated, so the MH algorithm is defined only in 
finite dimensions \cite{Stuart10} and degenerates under 
mesh refinement.  
However, a small modification to the classical random walk proposal 
\begin{equation}
\cQ(u_n,\cdot) = \cN(\sqrt{1-\beta^2} (u_n-m)+m, \beta^2 C)
\label{pcnprop}
\end{equation}
%This algorithm 
%of the original random walk proposal,
yields an algorithm which satisfies $\nu^T \ll \nu$
since $\cQ$ and $\mu_0$ are in detailed balance.  
This algorithm is therefore defined on function space, 
and it is referred to as pCN in \cite{CRSW12}.
The key point here is that the posterior has a density with respect 
to the infinite-dimensional prior Gaussian measure $\mu_0$
and so designing a proposal such as the one above
which is reversible with respect 
to this measure leads to the necessary condition 
$\nu^T \ll \nu$.
The condition that $\nu^T \ll \nu$ for algorithms on
function space ensures that the methods are well-behaved
with respect to mesh refinement; only such
methods are considered in this paper (see 
\cite{CRSW12} for more details on that point). 
The choice of proposal $\cQ$ which satisfies detailed balance 
with the prior $\mu_0$ is natural and elegant because it leads
to acceptance probability $\alpha(u,v)$ which depends only 
on the ratio of the likelihood evaluated at its arguments
\begin{equation}
\alpha(u,v) =  {\rm min} 
\left \{1, e^{\Phi(u)-\Phi(v)} \right \}.
\label{acceptance_pcn}
\end{equation}

\subsection{Correlations}
\label{correlation}

Let $R: H \rightarrow \bbR$, and 
suppose we would like to estimate the integral $\bbE (R) = \int_H R(u) \mu(du)$ 
%for some random quantity $R$ 
%(over $\bbR^M$ for simplicity) 
by $N$ samples $R_n=f(u_n)$, where 
$u_n \sim \mu$. 
Denoting this estimate by $\bar{R}$, we have
$$
\bar{R} = \frac{1}{N} \sum_{n=1}^N R_n.
$$
Then $\bbE(\bar{R}) = \bbE(R)$, so the estimator is without bias.
Furthermore, we denote the %$M\times M$ 
variance of $R$ by 
$$\bbV(R) = \bbE[ (R-\bbE(R))(R-\bbE(R))].$$
So, we also have that
$$\bbV(\bar{R}) = \bbE[ (\bar{R}-\bbE(R))(\bar{R}-\bbE(R))].$$
%=\bbE[ R\otimes
%(R-\bbE(R))] =\bbE[ (R-\bbE(R)) \otimes R] = \bbE[R \otimes R]-\bbE(R)
%\otimes \bbE(R)$.  
%
%Without loss of generality, assume 
We restrict attention to the case $\bbE(R)=0$, for 
ease of exposition only.  
Assume also that the samples are correlated in such a way that 
$$\rho_n = \bbE(R_m R_{m+n})/ \bbE(R_m R_{m})
~~ \forall m,n.$$
For example, such samples arise from a sample path of a 
homogeneous Markov chain, such as in MCMC.  Then
\begin{eqnarray}
\begin{array}{lll}
\bbV(\bar{R}) &=& \bbE[ \bar{R} \bar{R} ] \\
&=& (1/N^2) \left ( \sum_{n=1}^N R_n R_n + \sum_{n,m=1, n \neq m}^N
R_n R_m \right )\\
&\approx & (1/N) \bbV(R) (1+ 2 \sum_{n=1}^N \rho_n ) \\
& = & \frac{(1+ 2\theta)} {N} \bbV(R), 
\end{array}
\end{eqnarray}
where we assumed
$\rho_n \approx 0$ for all $n>N^*$ with $N^* \ll N$,
%the approximation is valid if $\rho_n \rightarrow 0$ and $N$ is
%large, 
and we denoted $\theta = \sum_{n=1}^N \rho_n$.  If the $R_n$
are independent then $\theta=0$ and we recover constant prefactor
$1$ to $\bbV(R)/N$.
Otherwise, the constant prefactor reduces the 
effective number of samples with respect to the 
case of independent samples.
It is therefore informative to define the {\it effective sample
  size} by
$N_{eff} = N /(1+2\theta)$.  Phrased another way, one 
needs $1+2\theta$ samples in order to get another 
approximately {\it independent} sample.  Hence, $\rho_n$ and $\theta$ 
are referred to as the autocorrelation function and the 
integrated autocorrelation, respectively.  The general case 
is dealt with similarly, working elementwise.
In this case, the effective sample size is given
by the minimum over effective sample sizes of individual degrees 
of freedom, strictly speaking.
% Strictly speaking 
% the effective sample size is given by $N_{eff}={\rm min}_n {\bf
%   N}^n_{eff}$ where ${\bf N}^n_{eff} = [{\bf N}^1_{eff}, {\bf
%   N}^2_{eff}, \dots]$ (assuming the space is countable).

From the above discussion, it is clear that the performance of an MH 
algorithm in approximating integrals with respect to a distribution 
is limited by the correlation between the samples it generates.  
The rest of the paper will concern this point about correlations.

\section{Proposal Chains for MCMC}
\label{proposal}

In this section, we explore some drawbacks of the function-space 
sampling algorithm defined in the previous section.  We will first 
motivate the need for modification with a simple one dimensional example.
Then, we describe how this same idea lifts into higher dimensions.

\subsection{Principle in 1d}
\label{principle}

\begin{figure}
 \includegraphics[width=0.5\textwidth]{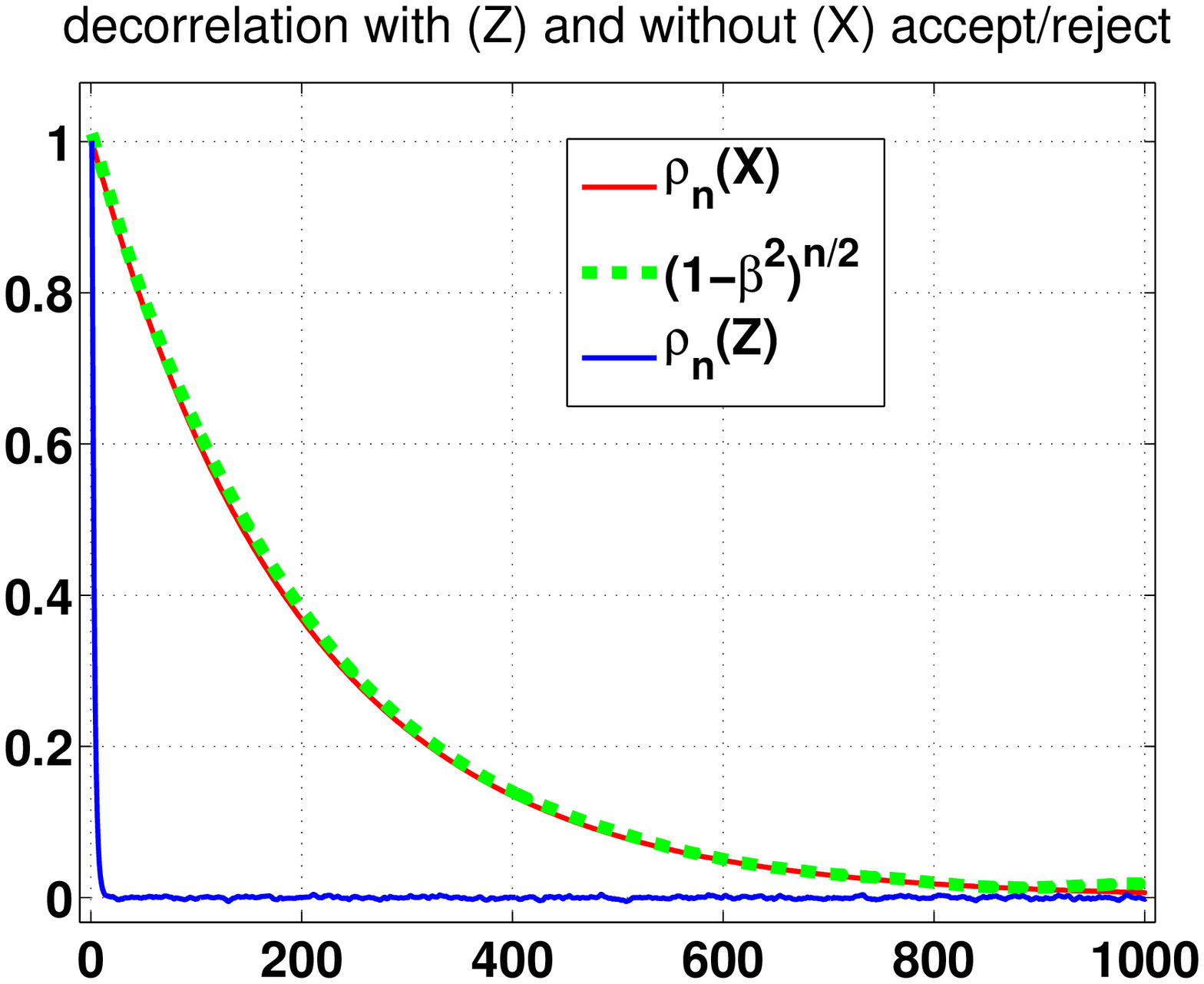}
 \includegraphics[width=0.5\textwidth]{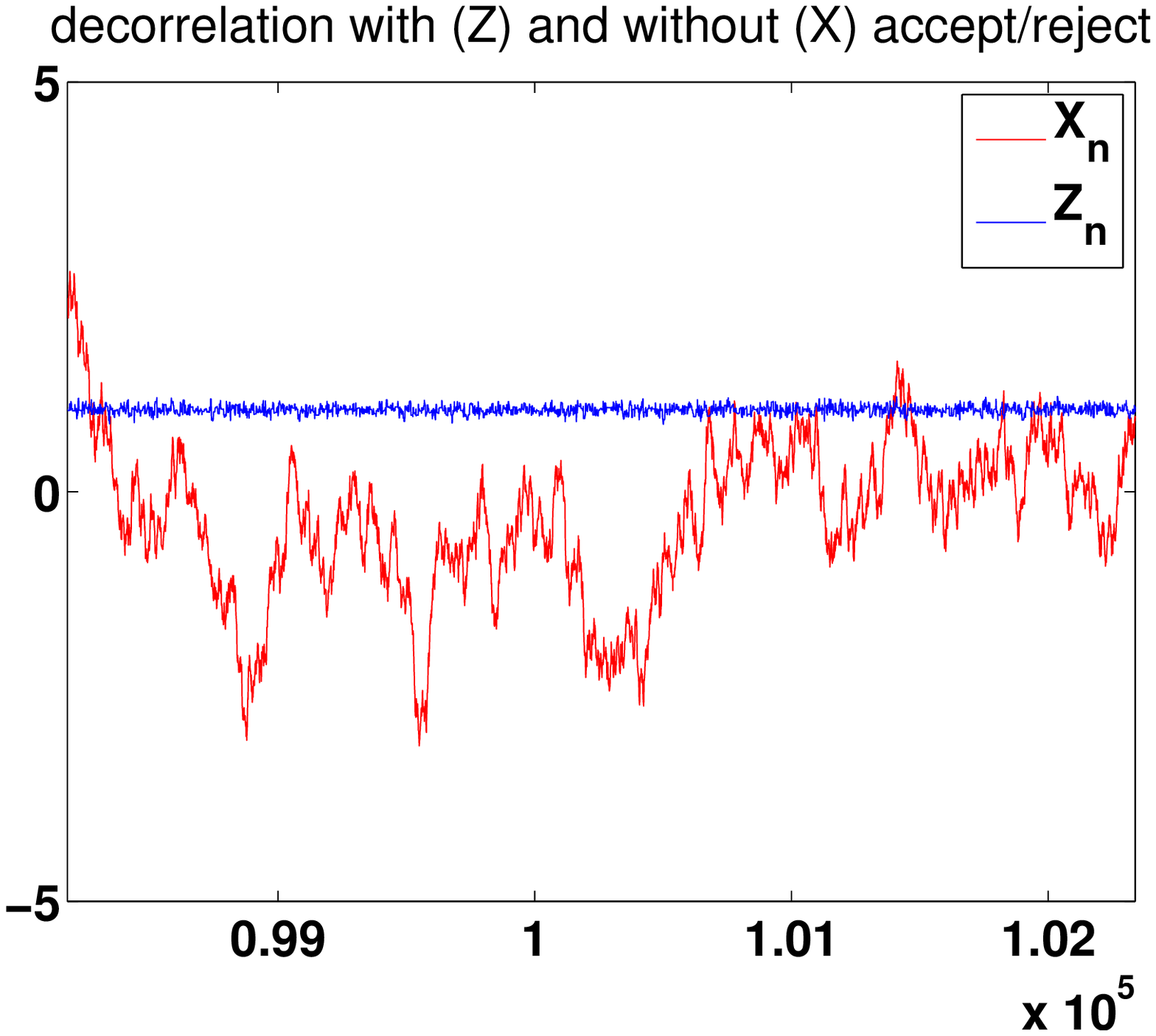}
 \caption{Illustration of slow decorrelation in proposal chain $X$, and
fast decorrelation in the informed chain $Z$. Left plot shows 
autocorrelations, as compared also to the analytical autocorrelation
of $X$.  Right plot shows an interval of the evolution over 
which $X$ has barely decorrelated, but $Z$
has clearly decorrelated within a small sub-interval.
}
 \label{mixing}
\end{figure}

As an example, consider the following Markov chain
$$
X_{n+1} = (1-\beta^2)^{1/2} X_n + \beta W_n
$$
where $X_0,W_n \sim \cN(0,\sigma^2)$ i.i.d. for all $n$ and $\beta<1$.
So, $\bbE X_n = 0$ for all $n$, and 
$$X_n = (1-\beta^2)^{n/2}X_0 +
\beta \sum_{k=0}^{n-1} (1-\beta^2)^{(n-1-k)/2} W_k.$$  Therefore,
$\rho_n = \bbE(X_0 X_n) / \bbE(X_0 X_0) = (1-\beta^2)^{n/2} \approx \exp(-n\beta^2/2)$,
and $\theta=\sum_{n=1}^N \rho_n \approx 2/\beta^2.$  
Notice also that this is independent of $\sigma$.

Now, we run an experiment.  Let $\phi(z) = (1/2\gamma^2) (z - 1)^2$,
with $\gamma=0.01$ and $\sigma=1$. 
Consider the following Metropolis-Hastings chain $Z$ 
with proposal given by $X$: 
%\begin{eqnarray}
\begin{equation}
\begin{array}{c}
%\centering
%\nonumber
Z^* = (1-\beta^2)^{1/2} Z_n + \beta W_n \\
Z_{n+1} \sim \alpha \delta(Z^*-~\cdot) + (1-\alpha) \delta(Z_n-~\cdot), \quad 
\alpha = 1 \wedge \exp(\phi(Z_n)-\phi(Z^*)).
%\centering
\label{example1d}
\end{array}
\end{equation}
%\end{eqnarray}
The above chain samples the 
posterior distribution with observation 
$y=z+\cN(0,\gamma^2)=1$ and prior $Z \sim \cN(0,1)$.
We will denote the chains by $X=\{X_n\}$ and $Z=\{Z_n\}$
Figure \ref{mixing} illustrates the decorrelation time and evolution 
of the proposal chain, $X$, and the informed chain $Z$ for a given
choice of $\beta$ which yields reasonable acceptance probability.  
The speedup is astonishing,
and indeed counterintuitive: one only accepts or rejects
moves from the proposal chain, so intuition may suggest that  
the informed chain would have larger autocorrelation.  
But actually, if the 
distribution is dominated by the likelihood, then the integrated autocorrelation
is determined by the accept/reject, and is indeed {\it smaller} 
than that of the proposal.  
On the other hand, if the distribution is 
dominated by the prior, then the integrated autocorrelation is dictated by the
proposal chain and is then increased by the acceptance probability.  
If the state-space were two dimensional, with 
an observation only in the first dimension, then for a given choice
of scalar $\beta$ the chain will mix like $Z$ in the observed 
component and like $X$ in the unobserved component.  
The same idea extends to higher dimensions, motivating
the use of direction-dependent proposal step-sizes.

\begin{figure}
 \includegraphics[width=5cm,height=5cm]{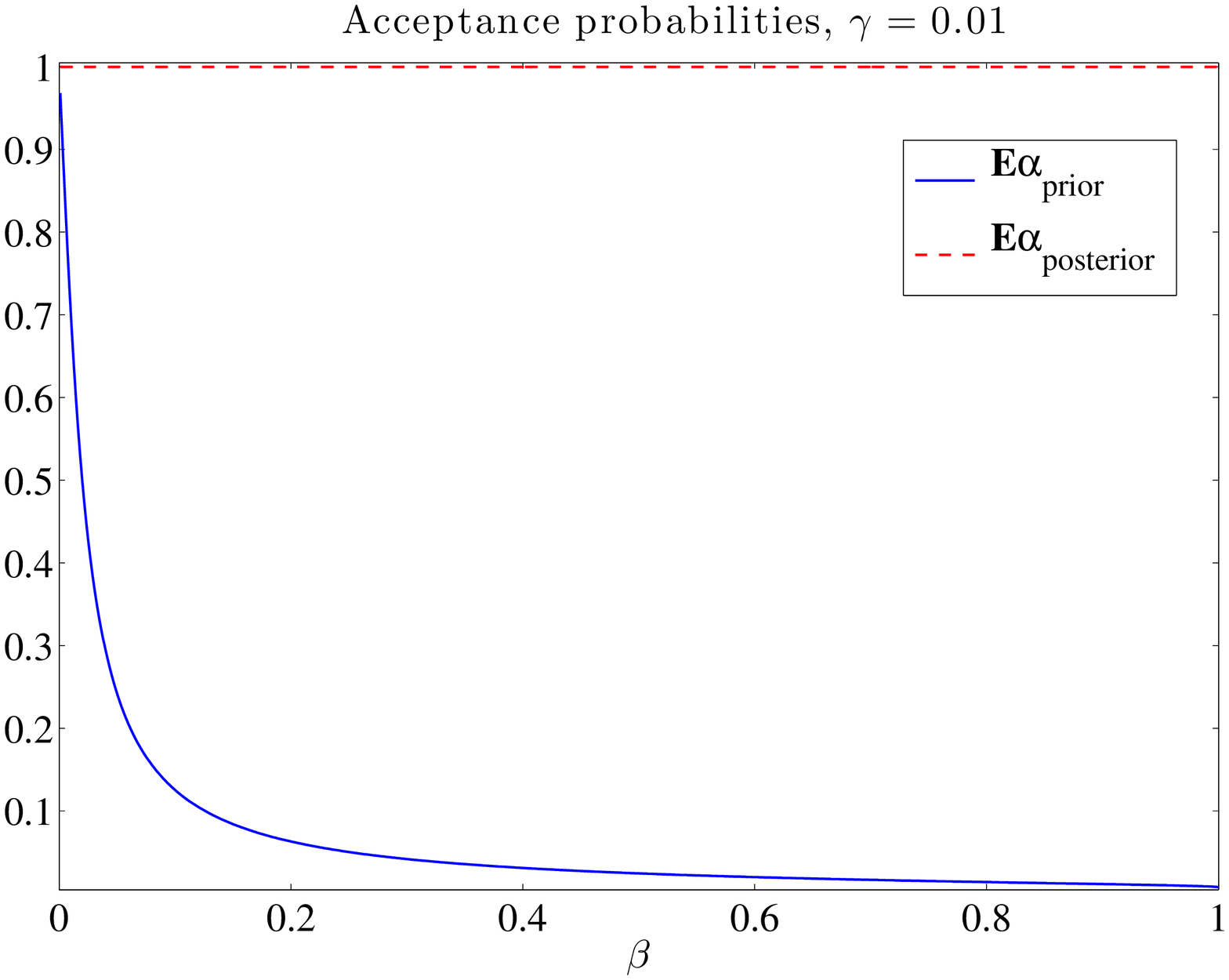}
 \includegraphics[width=5.4cm,height=5cm]{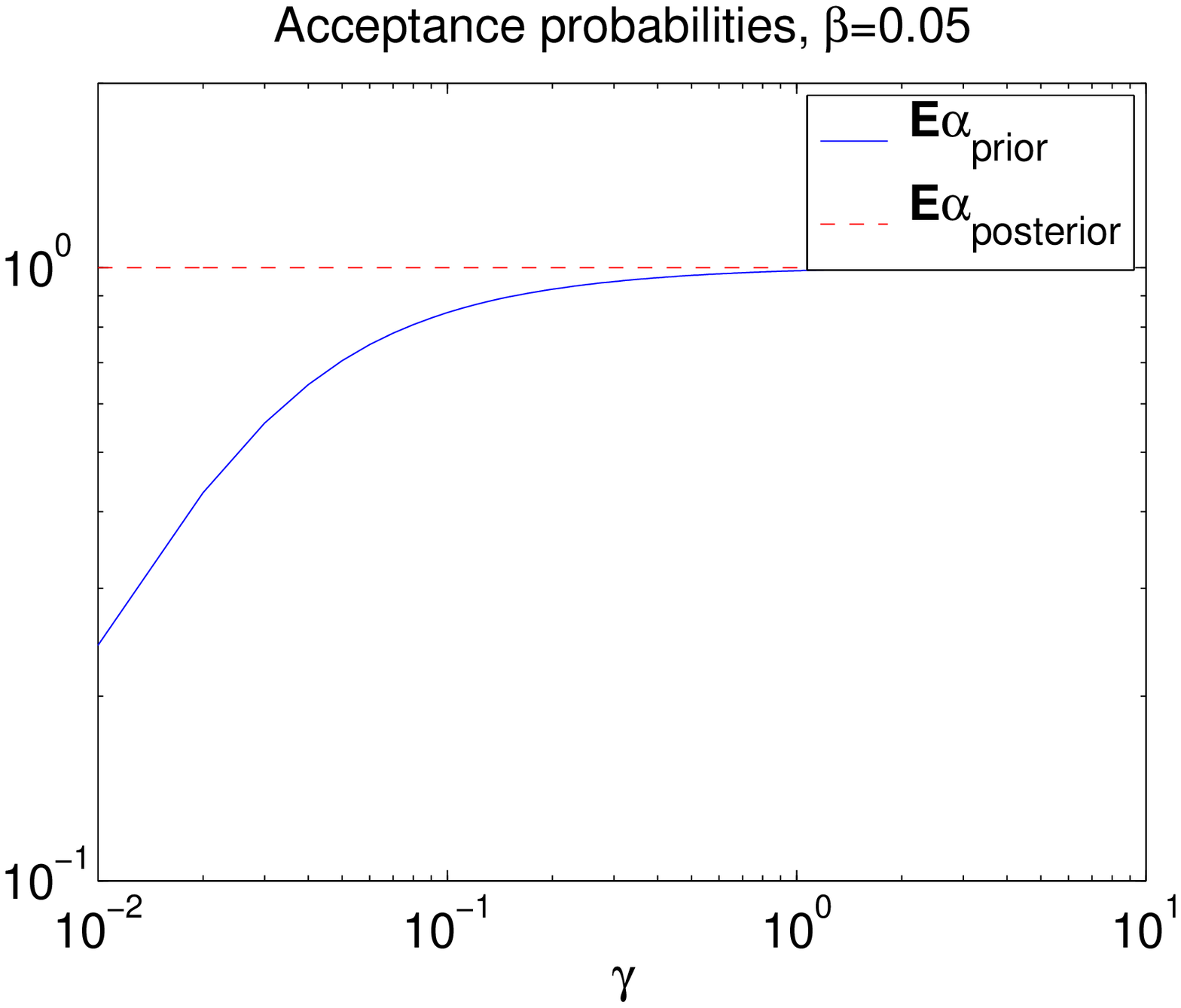} \\
 \includegraphics[width=5cm,height=5cm]{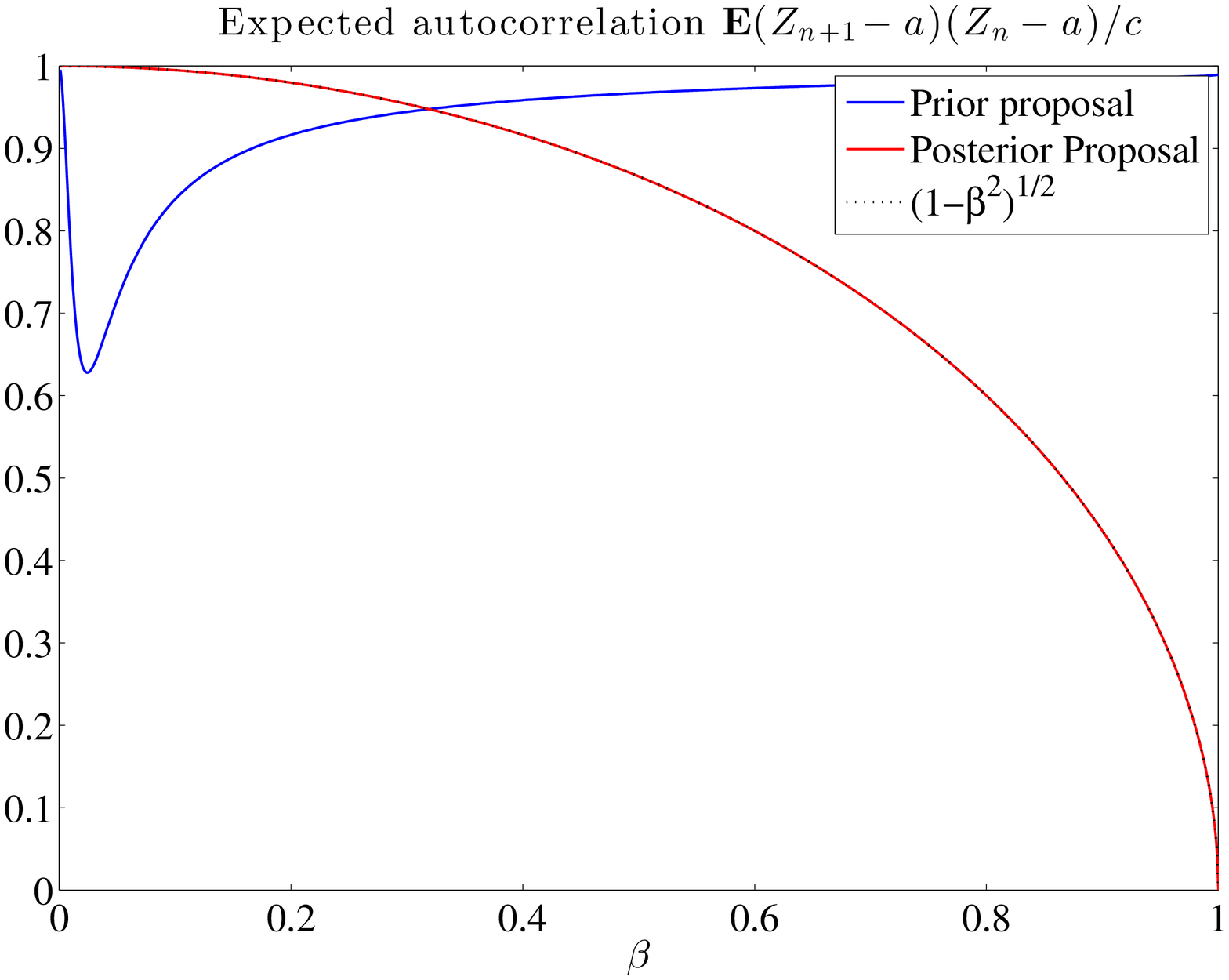}
 \includegraphics[width=5cm,height=5cm]{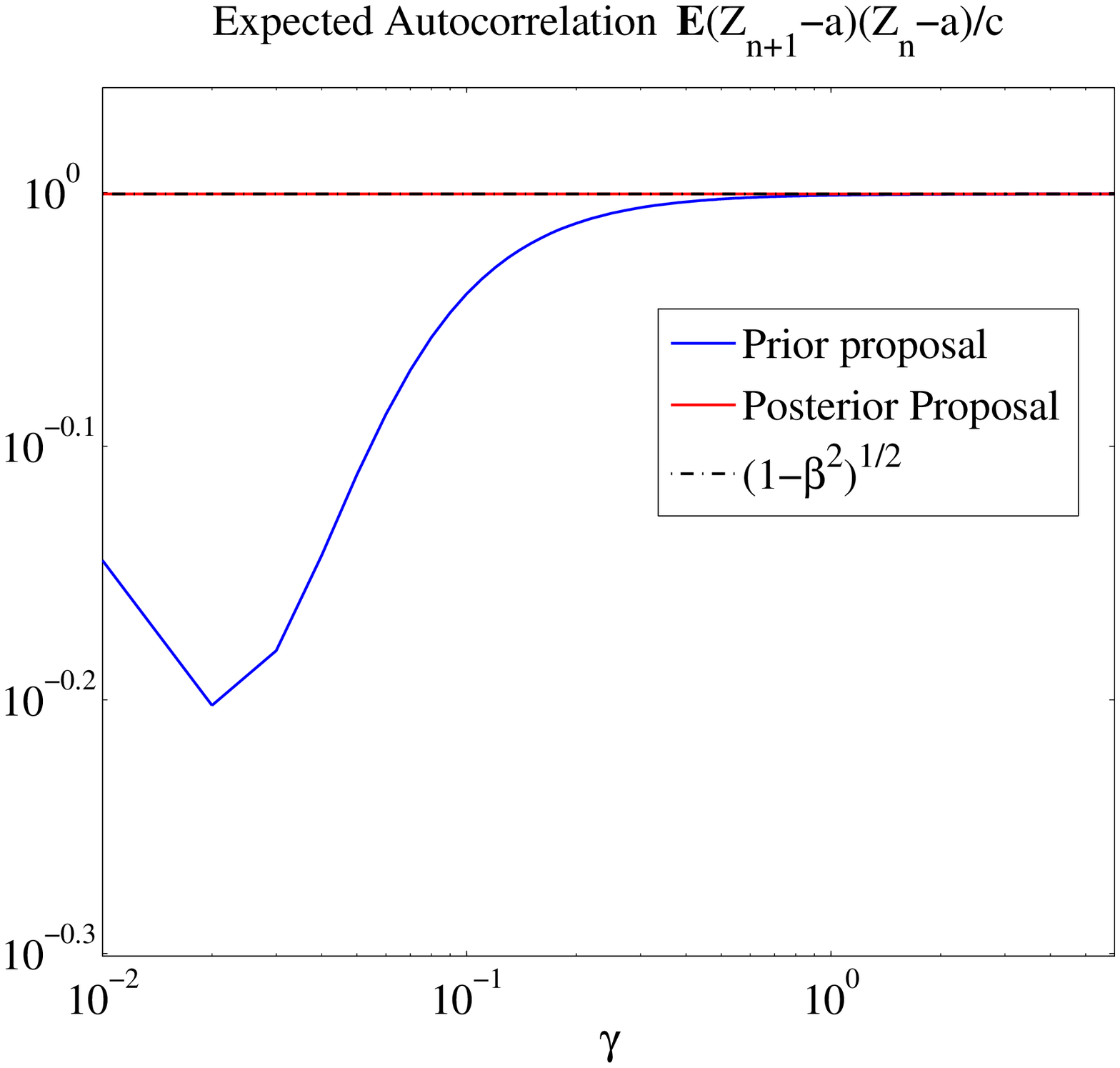} \\
 \caption{Expected acceptance probability (top) and lag 1
   autocorrelation (bottom) over $\beta$ with $\gamma=0.01$ (left) and
 over $\gamma$ with $\beta=0.05$ fixed (right), for proposals which
 are reversible with respect to the prior (blue) and the posterior
 (red dashed).  Also shown at the bottom is the lag 1 autocorrelation of 
the proposal chain, $\sqrt{1-\beta^2}$, in black dash-dotted.}
 \label{analytical}
\end{figure}

We now digress momentarily to explain this counterintuitive 
phenomenon in more detail for the simple analytically solvable example
above.  This is crucial to understanding the methods which will be
developed in subsequent sections.  In the above example, the posterior
distribution is given by $\cN(a,c)$, where $a=1/(\gamma^2+1)$ and 
$c=\gamma^2/(1+\gamma^2)$.  Assuming that $Z_n \sim \cN(a,c)$,
we know from Eqs. \ref{example1d} 
that the correlation between subsequent steps is given by the 
following analytically known, yet horrendous, integral
\begin{equation}
\begin{array}{lll}
\frac{1}{c}\bbE[(Z_{n+1}-a)(Z_n-a)] &=&\frac{1}{2\pi c\sqrt{c}}
 \int_{\bbR^2} [ z^*-a)(z-a)
\alpha(z,z^*) + \\
&& (z-a)^2 (1-\alpha(z,z^*)) ]
e^{-[\frac{(z-a)^2}{2c} + \frac{1}{2}w^2]} dz dw,
\end{array}
\label{lag1_exp}
\end{equation}
where $z^*=\sqrt{1-\beta^2}z+\beta w$.
It is instructive to also consider
in Eq. \eqref{example1d} the alternative 
proposal which keeps the posterior invariant:
\begin{equation}
Z^* = (1-\beta^2)^{1/2} (Z_n-a)+ a + \sqrt{c}\beta W_n,
\label{posterior_prop}
\end{equation}
in which case
$z^*=\sqrt{1-\beta^2}(z-a)+a+\sqrt{c}\beta w$ in Eq. \eqref{lag1_exp}
and the acceptance  \eqref{acceptance_pcn} is modified accordingly.
Indeed, it turns out that $\frac{1}{c}\bbE[(Z_{n+1}-a)(Z_n-a)] \leq
\sqrt{1-\beta^2}$, with equality in the limit of $\bbE\alpha \rightarrow 1$.
See also \cite{GGR97}.  We compute this integral using a Riemann sum
and plot the corresponding values of $\alpha$ and 
$\frac{1}{c}\bbE[(Z_{n+1}-a)(Z_n-a)]$ for varying $\beta$ with $\gamma$
fixed, and varying $\gamma$ with $\beta$ fixed in
Fig. \ref{analytical}.  In the top left panel, we see that for
proposal \eqref{example1d} with $\gamma=0.01$ the
acceptance decreases from 1 to 0 as $\beta$ ranges from 0 to 1.  In
the bottom left panel below this we can see that the minimum lag 1
autocorrelation corresponds to the acceptance probability 
$\bbE\alpha\approx 0.5$; c.f. the well-known 
optimal scaling results of \cite{GGR97}.  For the proposal
\eqref{posterior_prop}, the acceptance is 1 for all $\beta$.
It is clear from the bottom left
plot that in this example we should use \eqref{posterior_prop} with 
$\beta=1$ and do independence sampling on the known posterior.  
This obvious result in this idealistic example is instructive,
however.  It is not possible to do independence sampling
on the posterior for non-Gaussian posteriors (or we would be done)
and even independence sampling on the Gaussian approximation to 
the posterior may lead to unacceptably small acceptance probability.
Indeed we can see that the prior proposal can perform as well as or
better than the
posterior proposal for appropriate $\beta<0.75$, a step-size that is 
probably much too large in a Gaussian proposal whenever the 
target is non-Gaussian.  This makes it
clear that the crucial thing is to choose the appropriate scale of the 
proposal steps, 
rather than naively trying to match the target with the proposal 
(although this principle does hold for the case presented above in 
which the proposal keeps the target posterior invariant, and in general
this does lead to reasonably effective proposals).
From the top right panel, we can see that the acceptance probability
for the posterior proposal is again 1 for all $\gamma$ for a
fixed $\beta=0.05$, while the autocorrelation (bottom right) also 
remains approximately $\sqrt{1-\beta^2}$ (which is quite large in this
case).  However, as this choice of $\beta$ is tuned to the prior
proposal for $\gamma=0.01$, 
we see that the acceptance probability for prior proposal  
increases from $0.25$
to almost 1 as $\gamma$ ranges from $0.01$ to 5.  The corresponding
lag 1 autocorrelation ranges from $0.63$ to approximately
$\sqrt{1-\beta^2}$ as $\gamma$ increases and the posterior approaches 
the prior.

\subsection{Higher dimensions}
\label{higherd}

Recall that in one dimension if the 
distribution is determined by the likelihood then the integrated
autocorrelation is close to 1, 
and if the distribution is determined by the prior 
then the integrated autocorrelation is dictated by the
proposal chain.  Suppose that some of the space is dominated by the prior,
and some is dominated by the likelihood.  If the observations have
small variance, then the step-size $\beta$ in the chain will have to 
be small in order to accommodate a reasonable acceptance probability.
But, the part of the space which is dominated by the prior will 
have $O(1/\beta^2)$ autocorrelation, as dictated by the proposal chain. 

Now consider the Bayesian inverse problem on a Hilbert 
space $H$.
Let $\{\varphi_j\}_{j=1}^\infty$ denote an orthonormal 
basis for the space $H$.  
Let $P$ project onto a finite basis $\{\varphi_j\}_{j=1}^J$ and $Q=I-P$.
Suppose we wish to sample from the posterior distribution
$\mu$ with prior distribution $\mu_0$, 
where $d(\mu(Q \cdot), \mu_0(Q \cdot)) \ll \epsilon$ for some 
$\epsilon \ll 1$ and some distance function 
$d(\cdot,\cdot)$ between probability
measures (for example Hellinger metric or Kullback-Leibler distance).
Thus $P$ projects onto the 
``important part'' of the space, in the sense that the observations
only inform the projection of the posterior distribution on this
part of the space.  Suppose we use Metropolis-Hastings to 
sample from posterior $\mu$ with prior $\mu_0=\cN(0,C)$.
Let 
$u_0,W_n \sim \cN(0,C)$, so that the pCN chain 
\begin{equation}
u_{n+1} = (1-\beta^2)^{1/2} u_n + \beta W_n
\label{proposali}
\end{equation}
leaves the prior invariant.  Without loss of generality, we
represent this chain in the basis in which $C$ is diagonal, so
the proposal for each degree of freedom 
is an independent version of the above and 
clearly %by the above argument each direction 
has decorrelation time $1+2\theta \approx 4/\beta^2$.
The distribution $\mu \approx \mu_0$ on $QH$, so when we
impose accept/reject using the above chain as proposal,
we are sampling from $\mu(Q \cdot)$ with decorrelation time 
$(1+2\theta)/\bbE\alpha$ where $\bbE\alpha$ is the expected 
acceptance probability. 
But, we may as well sample independently from the known 
distribution $\mu_0(Q \cdot)$ in this part of the space.

Following the reasoning above, one 
can consider modifying the above proposal chain to sample
from the pCN chain \eqref{proposali} on $PH$ and sample independently
from the prior on $QH$.  This corresponds to setting the chain above to
\begin{equation}
u_{n+1} = (1-\beta^2)^{1/2} P u_n + (\beta P + Q) W_n.
\label{prop1}
\end{equation}
Our first and simplest variant of weighted proposal follows directly 
from this.  In the next section we will consider a more
elaborate extension of this proposal using curvature information.

\section{Operator weighted proposal}
\label{operator}

Inspired by the discussion in the preceding section, 
we let $B_n$ be an {\it operator} weighting different
directions differently, and define the proposal chain to be
\begin{equation}
u_{n+1} = \sqrt{B_n} u_n + \sqrt{I-B_n} W_n.
\label{operator_beta}
\end{equation}
with $u_0,W_n \sim \cN(0,I)$.  Notice that this proposal preserves the 
prior in the sense that $u_n \sim \cN(0,I)~~ \forall n$.  
This form of proposal for $B_n=B$ constant in $n$ 
can be derived as a generalization of the pCN 
proposal from \cite{CRSW12}, for some general 
pre-conditioner $\cK$ in their notation. 
%, and may be referred to as generalized pCN.

Recall the form of distribution we consider, given by \eqref{density}
and \eqref{loglike}, %.  We will consider 
with prior %densities 
of the form
%Consider the Bayesian inverse problem in which 
$u \sim \cN(m,C)$.
%If the prior covariance $C \neq I$, then 
For simplicity of exposition, 
we will consider standard normal $\cN(0,I)$ priors, without
any loss of generality.  In the general case, a change of 
variables $u \rightarrow C^{1/2}u+m$ is imposed to normalize the prior,
such that a chain of the 
form \eqref{operator_beta} keeps the prior invariant.
The samples of the resulting MH chain are then
transformed back as $u \rightarrow C^{-1/2}(u-m)$ for inference.
%For simplicity of exposition in what follows, we will let $m=0$
%without any loss of generality.

For the moment, we revert to the case in which $H=\bbR^N$
in order to motivate the form of the proposals.
%and 
%$$
%y = g(u) + \eta,
%$$
%where $\eta \sim \cN(0,\Gamma)$.
In this case we have $\mu \ll \lambda$, where $\lambda$ is Lebesgue
measure, and after transformation
\begin{equation}
- {\rm log} \left [ \frac{d\mu}{d\lambda}(u)  \right ] = 
\Phi(u)+ \frac{1}{2}|u|^2 +K(y),
\label{logpost}
\end{equation}
%Changing variables $u \rightarrow C^{-1/2}u$, %and neglecting the
%constant in $u$
%we have
%$$
%I(u) = \frac{1}{2\gamma^2} |y-G(C^{1/2}u)|^2 + \frac{1}{2}|u|^2 + K(y).
%$$
with $K(y) = \int_H \exp\{-[\Phi(u) + \frac{1}{2}|u|^2]\} d\lambda$.
At a given point $w$ the local curvature is given by the Hessian
of this functional,
which can be approximated by the positive definite operator
\begin{equation}
\frac{1}{\gamma^2} DG(w)^* DG(w) + I,
\label{structure}
\end{equation}
where $DG(w)$ denotes the derivative of $G$ with respect to 
$u$ evaluated at $w$ \footnote{The neglected term is
  $\sum_i D^2G(w)_{i,jk}(y-G(w))_i$, which is close to $0$ for example 
if the distribution is concentrated close to the data or if
$G$ is close to quadratic at $w$.}.
If $\{\lambda_k,\phi_k\}$ are eigenpairs
associated with this operator, then:
%the first term in Eq. (\ref{structure}):
\begin{itemize}
\item if $\lambda_k \gg 1$, then the distribution 
is dominated by the data  
likelihood in the direction $\phi_k$;
\item if $\lambda_k \approx 1$, then the distribution is dominated by the
  prior in the direction $\phi_k$.
\end{itemize}
%The covariance of a local Gaussian approximation 
%is determined by the inverse of \eqref{structure}.  
The direction of the largest curvature at a given point will determine 
the smallest scale of the probability distribution at that point.  If $\lambda_1$
is the largest eigenvalue of \eqref{structure}, 
then the smallest local scale of the density will be $\sqrt{1/\lambda_1}$.
This scale will directly determine the 
step-size necessary to obtain a reasonable acceptance probability 
within the MH algorithm.  But,
such a small step-size is only necessary in the direction $\phi_1$.  So, rather
than prescribing this as the step-size in the remaining directions, we
would like to {\it rescale} them according to the operator given in
\eqref{structure}.  

%The log-density is not 
There is no log-density with respect to Lebesgue measure in
function space, but the above operator \eqref{structure} is
still well-defined and indeed its minimizer gives the point of
maximum probability over small sets \cite{ourmap}.
This functional will be the basis for 
choosing $B_n$ in \eqref{operator_beta}. 
Based on the above intuition, one can devise a whole 
hierarchy of effective such operators.   
In the simplest instance, one can allow for $B_n$ constant, either
during the duration of the algorithm, after a transient adaptive
phase, or even for some lag-time where it is only updated every so
often.  
In this manuscript we consider only constant $B$, 
%(after a
%transient adaptive phase which will not be discussed in more detail), 
and we consider three simple levels of complexity.
The rigorous theory underlying the algorithms will be deferred to
future work, although we note that the case of constant $B$ 
is easier to handle theoretically, since one does not need to 
consider the conditions for ergodicity of adaptive MCMC 
\cite{roberts2007coupling}. 
%For example, the next level of complexity involves letting
%${\bf \beta} = (1-\beta^2)^{1/2} {\bf \beta}(k)$, where 
%${\bf \beta}(k) = {\bf 1}_{k<k_1} + (k-k_1)/(k_1-k_2) {\bf 1}_{k_1<k<k_2}$.
%For the final level of complexity, we will focus on a particular form
%of Bayesian inverse problem.

The first level of complexity consists of letting 
\begin{equation}
B = (1-\beta^2) I. 
\label{original}
\end{equation}
The resulting method corresponds to the original pCN proposal,
and we denote this method by O.  
 
The second level of complexity consists of letting 
\begin{equation}
B_{k,j} = (1-\beta^2) {\bf 1}_{k<k_c} \delta_{k,j}, 
\label{characteristic}
\end{equation}
where $B=\{B_{k,j}\}_{k,j=1^\infty}$, $\delta_{k,j}$ is the Kronecker delta function, 
and $k_c \in (0,k_{\rm max}]$ and $\beta \in (0,1]$ are chosen during a transient adaptive 
phase, either together or one at a time, yielding \eqref{prop1}.
The simple idea is that the curvature is constant and prescribed by the prior
for sufficiently high frequencies, and this has already been motivated 
by the discussion in the previous section. 
The method resulting from this proposal will be denoted by C.

The third level of complexity consists of utilizing a relaxed version
of the Hessian for an arbitrary given point close to a mode of the 
distribution. \footnote{In the case of a multi-modal distribution, 
this would have to be updated at least each time the chain passes to a new 
peak of the distribution.  However, in the case of 
high-dimensional sampling, it is rare that one might expect to
sample more than a single peak within a single chain, since it is 
expensive to compute individual samples and many such samples 
are usually necessary to reach the first passage time.  Therefore, we
separate the issues of {\it finding a peak} of the distribution, where
gradient information is invaluable, from {\it exploring the peak}, 
where gradient information is expected to be less valuable.  On the
other hand, curvature information may be useful in the former, and is 
surely invaluable in the latter.}
%In the case that $G$ is linear, we know the optimal $B$, which is
%Consider $B$ in
%Eq. (\ref{operator_beta}) 
%given as
Let
\begin{eqnarray}
%\nonumber
%B &=& (1-\beta^2) (C^{1/2} DG(C^{1/2}w)^*DG(C^{1/2}w) C^{1/2} +
%\zeta \gamma^2I)^{-1} \\ && C^{1/2} DG(C^{1/2}w)^*DG(C^{1/2}w)
%C^{1/2},
B &=& (1-\beta^2) (DG(w)^*DG(w) +
\zeta \gamma^2I)^{-1} DG(w)^*DG(w),
% B &=& (1-\beta^2)^{1/2}/\lambda_1 
% C^{1/2} Dg(C^{1/2}w)^* \Gamma^{-1} Dg(C^{1/2}w) C^{1/2},
\label{optimalb}
\end{eqnarray}
where $\beta$ is the scalar tuning parameter 
as in the above cases, $\zeta \in (0,1]$ is another tuning parameter, 
and $w$ is any point. 
The method resulting from this proposal will be denoted by H. 
%For the nonlinear 
%case, this particular form will not be effective, and 
%we instead consider sampling a single mode by
%setting $B$ to be a relaxed version of the above, where 
%$\gamma \rightarrow \zeta \gamma$ with $\zeta<1$.  
%Furthermore, 
We consider a low rank approximation
of the resulting operator, since it is assumed to be compact.
%and this is relevant to a wide range of inverse problems.  
Notice that for this choice of $B$, the covariance of the search
direction is given by  
\begin{eqnarray}
\nonumber
% I-B &=& (C^{1/2} DG(C^{1/2}w)^* DG(C^{1/2}w) C^{1/2}/(\zeta \gamma^2) + I)^{-1} 
% + \\
% \nonumber
% && \beta^2(C^{1/2} DG(C^{1/2}w)^* DG(C^{1/2}w) C^{1/2}+\zeta \gamma^2I)^{-1}
% \\
% && C^{1/2} DG(C^{1/2}w)^*DG(C^{1/2}w) C^{1/2}.
I-B &=& (DG(w)^* DG(w)/(\zeta \gamma^2) + I)^{-1} 
+ \\ 
&& \beta^2(DG(w)^* DG(w)+\zeta \gamma^2I)^{-1}
 DG(w)^*DG(w).
\label{search_dir}
\end{eqnarray}
For linear $G$ we let $\zeta=1$, and 
the right-hand side of the above approaches the 
covariance of the posterior distribution as $\beta \rightarrow 0$,
giving exactly the correct rescaling for the search direction while
still keeping the prior invariant.  
The linear case is rather special,
since we could just do independence sampling of the posterior, 
but it provides a good benchmark.  In practice, adaptation of $\beta$
in this case leads to small, but non-zero, $\beta$.
For nonlinear $G$ we choose $\zeta < 1$, and 
the scaling of the search direction is 
commensurate with the local curvature 
except with increased weight 
on the log-likelihood component, 
given by the first term of \eqref{search_dir}.
The second term is a perturbation which is larger 
for the (changing) directions dictated by the Hessian of the 
log-likelihood and approaches zero for 
directions which are dictated by the prior, in which the
curvature is constant.  Provided $\zeta$ is chosen small 
enough, and the curvature does not change too much, 
the step-size of a given move will not exceed the scale of the 
distribution.  Furthermore, the step-size will 
interpolate appropriately between the 
direction of the smallest scale and the directions
which are determined by the prior, corresponding to 
scale 1.  We note that a whole array of different proposals 
are possible by choosing different $B_n$, 
and there may be better options since an 
exhaustive search has not been carried out.  
Investigation into various other proposals 
which incorporate the appropriate (re)scaling in different 
ways, however, suggests that their performance is similar 
\cite{TC}.   

Both methods C and H can be considered 
as splitting methods in which part of the space is targeted 
using the acceptance probability and independence sampling is 
done on the complement.  C splits the spectral 
space into a finite inner radius in which pCN is done, 
and its infinite-dimensional complement where independence 
sampling is done.  H more delicately splits the space into the 
finite dominant eigenspace of 
the Hessian of the likelihood $DG^*(w)DG(w)$ and its 
complement, and then furthermore considers an appropriate 
rescaling of the dominant eigenspace.  Notice that 
the complement space absorbs the infinite-dimensionality.
Such proposals which separate the space allow any general form of 
proposal on the finite-dimensional target subspace and maintain
the property that $\nu^T \ll \nu$ as long as the proposal is reversible
with respect to the prior on the infinite-dimensional complement 
subspace.  For example, a simple pCN proposal with another 
choice of $\beta$ can be used on this subspace.  This idea appears
in the forthcoming works \cite{TC,bckj}.

Note that one could instead incorporate this curvature information 
by devising a chain which keeps a Gaussian approximation to the
posterior invariant, similarly to \eqref{posterior_prop}.  For example, 
one can use the following proposal which preserves $\cN(\xi,\Xi)$: 
\begin{equation}
u_{n+1} = \Xi^{1/2} \sqrt{1-B_n} \Xi^{-1/2}(u_n-\xi) + \xi + (\Xi B_n)^{1/2} W_n,
\label{posterior_propd}
\end{equation}
with $u_0\sim \cN(\xi,\Xi)$ and $W_n \sim \cN(0,I)$. 
The curvature information can be incorporated for example 
by letting 
$$\xi= {\rm argmin}_u \left [\Phi(u) + \frac{1}{2}|u|^2\right ],$$
or some close by point, and 
$\Xi=(\frac{1}{\gamma^2} DG(\xi)^* DG(\xi) + I)^{-1}$.
In this case, it would be sensible to revert to the case of scalar
$B_n$ since the relevant scaling is already intrinsic in the form
of the Gaussian. 
Empirical statistics can also be incorporated in this way.  
If $DG(\xi)^* DG(\xi)$ is genuinely low-rank then equivalence 
$\nu^T \ll \nu$ is immediate.  If not, one must look more closely.
However, in this case if one imposes a low-rank approximation,
then equivalence is again immediate based on the discussion above.
The results in Sec. \ref{principle}, and in particular those presented 
in Fig. \ref{analytical} illustrate that this strategy should not be 
considered as preferable to using a proposal of the form
\eqref{operator_beta} for appropriate choice of $B_n$. 
These variations do however warrant further investigation.
We limit the current study to the case of proposals of the form
\eqref{operator_beta} because (i) these methods are defined on 
function space, following immediately from the reversibility with 
respect to the prior and the framework of \cite{Tie}, (ii)
the acceptance probability takes the simple and elegant form of 
\eqref{acceptance_pcn} which is easy to implement, and (iii) as
described in Sec. \ref{proposal}, it is 
not clear that a more elaborate kernel such as the one following 
from the proposal \eqref{posterior_propd} would yield better results.

\section{Numerical Results}
\label{numerics}

In this section, we will investigate the performance of the 
MH methods introduced
in the previous section on two prototypical examples.  First,
we briefly explore the inverse heat equation on the 
circle as an illustrative example.  
Next, we look in more depth at the more complicated problem of 
the inverse Navier-Stokes equation on the 2D torus.

\subsection{Heat Equation}
\label{heateq}

%!! Make like NSE section !!

Consider the one dimensional heat equation
\begin{eqnarray*}
\frac{\partial v}{\partial t} - \frac{\partial^2 v}{\partial x^2} &=& 0\\
v(t,0)=v(t,\pi)&=&0.
\end{eqnarray*}
We are interested in the inverse problem of recovering 
$u=v(0,x)$ from noisy observations of $v(1,x)$:
\begin{eqnarray*}
y&=&v(1,x)+\eta\\
&=&G u+\eta,
\end{eqnarray*}
where $G=\exp \{-A \}$, $A=(-\frac{d^2}{dx^2})$, 
$D(A)=H^2(\Omega)\cap H^1_0(\Omega)$ with $\Omega=(0,\pi)$, and
$\eta \sim N(0,\gamma^2 I).$
We let $C=10^4 A^{-1}$, and
$\gamma=1$. %to ensure we can clearly illustrate the point.

\begin{figure}
 \includegraphics[width=0.5\textwidth]{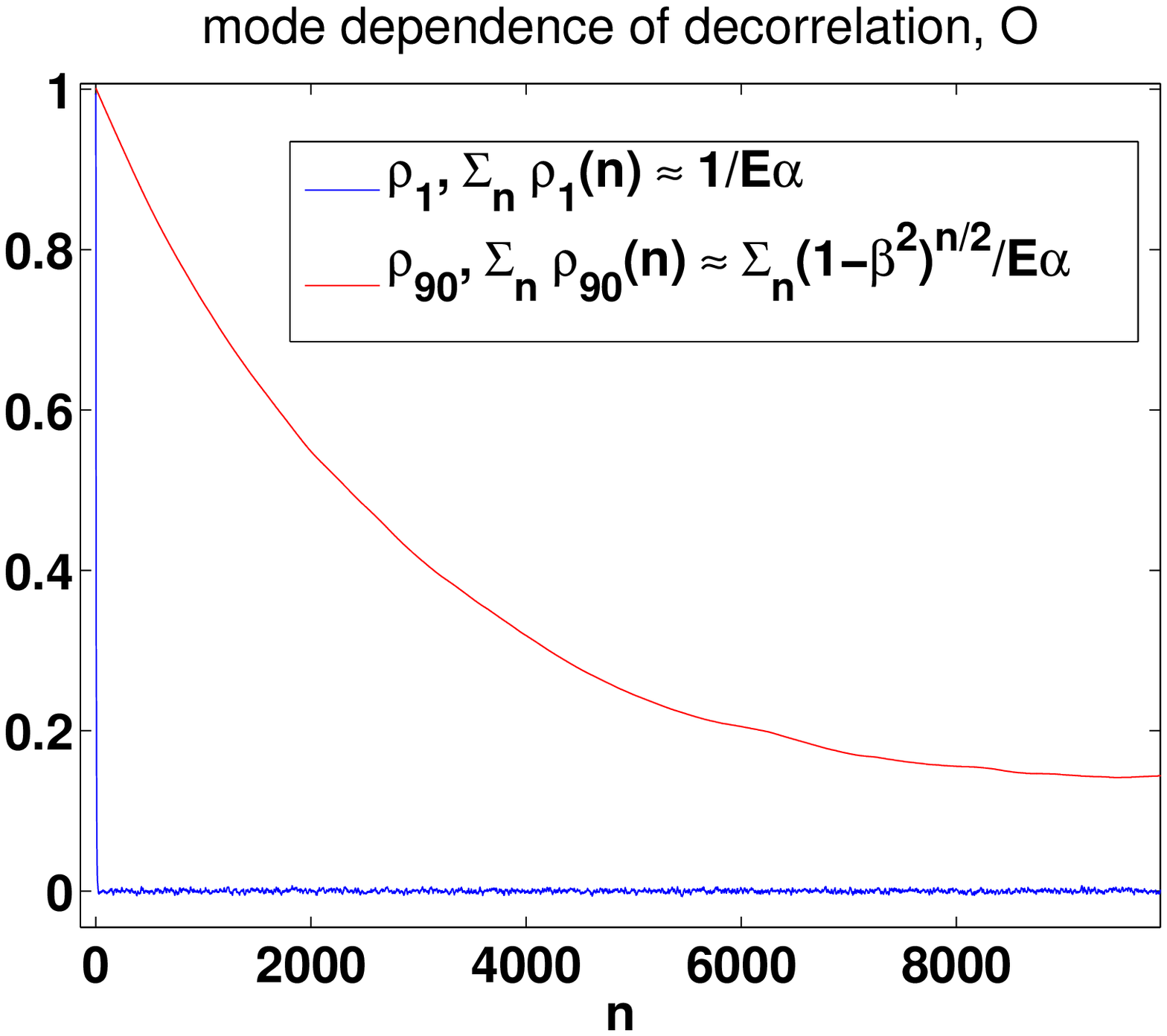}
 \includegraphics[width=0.5\textwidth]{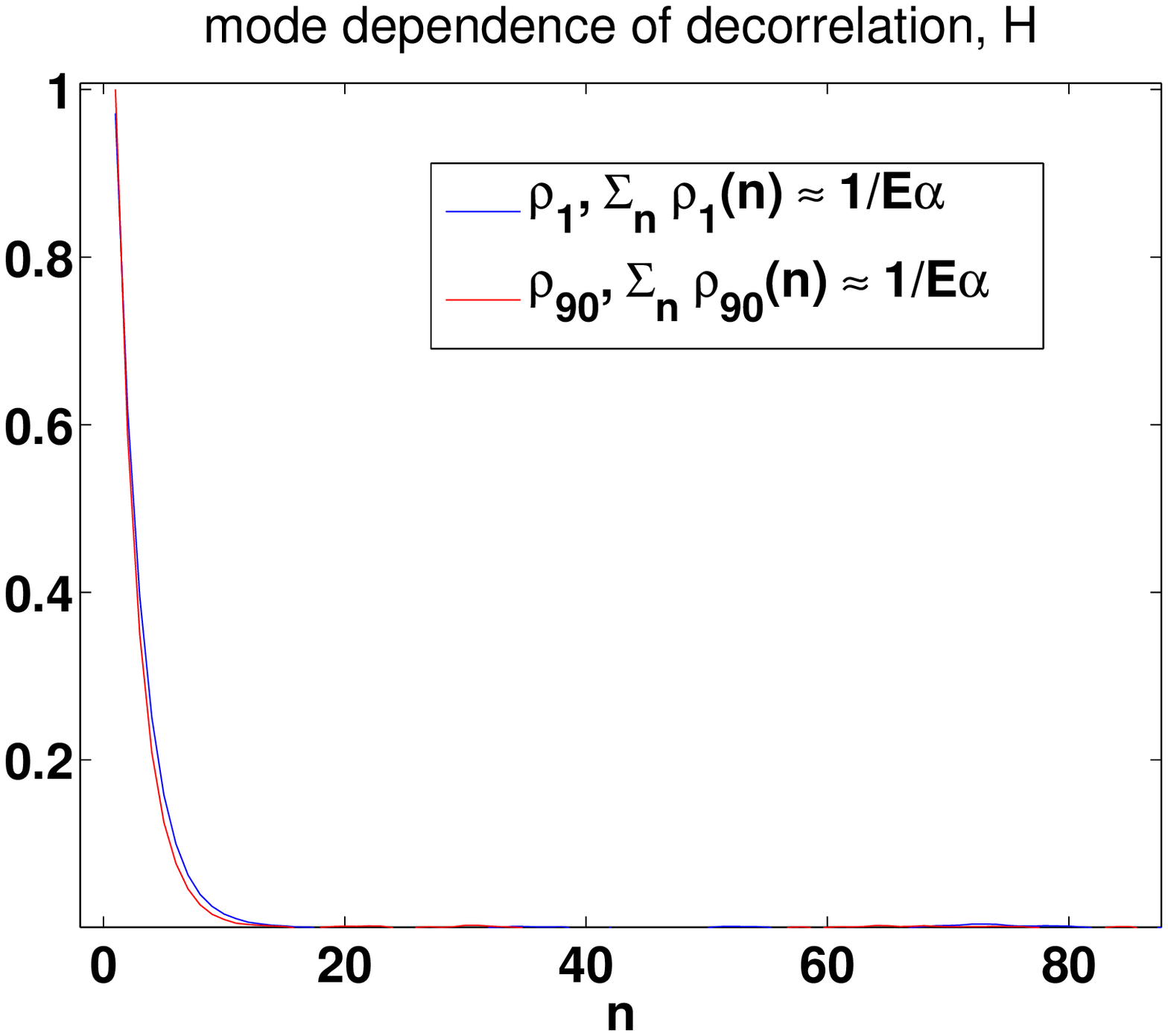}
\caption{Illustration of the autocorrelation of modes $u_1$ and
  $u_{90}$, in the case of O (left), and H (right).}
\label{comp_op}
\end{figure}

% Consider the Bayesian interpretation of the inverse heat
% equation on the circle given an observation at time 1.
% In this case $G(u) = L u = {\rm diag} \{ e^{-k^2} \} u$,
% where $u \in \bbR^{100}$ and $k \in [1,2,...,100]$.  This is
% the Fourier representation of the forward map of the heat equation 
% up to time 1 on the circle.  
The problem may be solved explicitly in the Fourier
sine basis, and the coefficients of the 
mean $m_k$ and variance $V_{k}$
of the posterior distribution can be 
represented in terms of the coefficients of the
data $y_k$ and the prior $C_{k,k}=10^4 k^{-2}$:
\begin{eqnarray}
\nonumber
m_k &=& (e^{-2k^2}+10^4 k^{2})^{-1} e^{-k^2} y_k, \\ 
V_k &=& (e^{-2k^2}+10^4 k^{2})^{-1}.
\label{exact_heat}
\end{eqnarray}
The posterior distribution is strongly concentrated on the 
$u_1$ mode, in the sense that $m_1 \gg m_j$ and $V_1 \gg V_j$ 
for all $j>1$, and the discrepancy between $\Gamma$ and $C$ will
necessitate a very small scalar $\beta$.  For the case
of scalar weighting O, 
all modes will decorrelate slowly except for $u_1$.  See Fig. 
\ref{comp_op} for a comparison between scalar weighting O 
and curvature-based weighting H.  In this case, we will omit
the intermediate proposal which simply proposes independent 
samples outside a certain radius in the spectral domain.

\subsection{Navier-Stokes Equation}
\label{nse}

In this section, we consider
the inverse problem of determining the initial condition of 
Navier-Stokes equation on the two dimensional torus given noisy 
observations.
This problem is relevant to data assimilation 
applications in oceanography and meteorology.

Consider the 2D Navier-Stokes equation on the torus $\TT^{2} := [-1,1)
\times [-1,1)$ with periodic boundary conditions:
% \begin{subequations}
\begin{eqnarray*}
  \begin{array}{cccc}
    % \begin{align}
    \partial_{t}v - \nu \Delta v + v \cdot \nabla v + \nabla p &=& f
    & {\rm for~ all ~}
    % \text{for all }
    (x, t) \in \TT^{2} \times (0, \infty), \\
    \nabla \cdot v &=& 0 &{\rm for~ all~ }
    % \text{for all }
    (x, t) \in \TT^{2} \times (0, \infty), \\
    v &=& u &{\rm for~ all ~}
    % \text{for all }
    (x,t) \in \TT^{2} \times \{0\}.
    % \end{align}
  \end{array}
  \label{eq:NSE}
\end{eqnarray*}
% \end{subequations}
Here $v \colon \TT^{2} \times (0, \infty) \to \R^{2}$ is a
time-dependent vector field representing the velocity, $p \colon
\TT^{2} \times (0,\infty) \to \R$ is a time-dependent scalar field
representing the pressure, $f \colon \TT^{2} \to \R^{2}$ is a vector
field representing the forcing (which we assume to be time independent
for simplicity), and $\nu$ is the viscosity.
We are interested in the inverse problem of
determining the initial velocity field $u$ from
pointwise measurements of the velocity field at later times.
This is a model for the situation in weather forecasting
where observations of the atmosphere are used to improve
the initial condition used for forecasting.
For simplicity we assume that the initial velocity
field is divergence free and integrates to zero
over $\TT^2$, noting that this property will be preserved
in time.

Define
\begin{equation*}
  {\mathcal H}:= \left\{ {\rm {\,trigonometric\,polynomials\,}}
    u\colon \TT^2 \to {\mathbb R}^2\,\Bigl|\, \nabla \cdot u = 0, \,\int_{\TT^{2}} u(x) \, \rd x = 0 \right\}
\end{equation*}
% \]
and $H$ as the closure of ${\mathcal H}$ with respect to the
$(L^{2}(\TT^{2}))^{2}$ norm. We define $P\colon (L^{2}(\TT^{2}))^{2}
\to H$ to be the Leray-Helmholtz orthogonal projector (see
\cite{book:Robinson2001}).  Given $k = (k_{1}, k_{2})^{\mathrm{T}}$,
define $k^{\perp} := (k_{2}, -k_{1})^{\mathrm{T}}$. Then an
orthonormal basis for $H$ is given by $\psi_{k} \colon \R^{2} \to
\R^{2}$, where
\begin{equation*}
  \psi_{k} (x) := \frac{k^{\perp}}{|k|} \exp\Bigl( \pi i k \cdot
  x \Bigr)
\end{equation*}
for $k \in \Z^{2} \setminus \{0\}$.
Thus for $u \in H$ we may write
\begin{equation*}
  u = \sum_{k \in \Z^{2} \setminus \{0\}} u_{k}(t) \psi_{k}(x)
\end{equation*}
where, since $u$ is a real-valued function, we have the
reality constraint $u_{-k} = - \bar{u}_{k}$.
Using the Fourier decomposition of $u$, we define the
fractional Sobolev spaces
\begin{equation}\label{eq:Hs}
  H^{s}:= \Bigl\{ u \in H \Bigm| \sum_{k \in \Z^{2} \setminus \{0\}} (\pi^{2}\abs{k}^{2})^{s}\abs{u_{k}}^{2} < \infty \Bigr\}
\end{equation}
with the norm $\norm{u}_{s}:= \bigl(\sum_{k} (\pi^{2}\abs{k}^{2})^{s}
\abs{u_{k}}^{2}\bigr)^{1/2}$, where $s \in \R$.  If $A=-P\Delta$, the
Stokes' operator, then $H^s=D(A^{s/2})$.  We assume that $f \in H^s$
for some $s>0$.

Let $t_n=n h$, for $n=0, \dots, N$, and let $v_n \in \bbR^M$ be the
set of pointwise values of the velocity field given by
$\{v(x_m,t_n)\}_{m = 1}^{M/2}$.  Note that each $v_n$ depends on $u$
and define $G_n \colon H \to \bbR^{M}$ by $G_n(u)=v_n$.  We
let $\{ {\eta}_n \}_{n=1}^N$ be a set of random
variables in $\bbR^{M}$ which perturbs the points $\{v_n \}_{n =
  1}^N$ to generate the observations $\{ y_n \}_{n=1}^N$ 
in $\bbR^{M}$ given by
\begin{equation}\label{eqn:Observation}
  y_n := v_n + \gamma {\eta}_n, \quad n \in \{1,\ldots, N\}.
\end{equation}
We let $y=\{y_n\}_{n=1}^N$ denote the accumulated data up to time $T=Nh$,
with similar notation for $\eta$, and define $G\colon H \to
\bbR^{MN}$ by $G(u)= \bigl(G_1(u),\dots,G_N(u)\bigr)$.  We now
solve the inverse problem of finding $u$ from $y=G(u)+\eta$.  
We assume that the prior distribution on $u$ is a Gaussian
$\mu_0=N(0,C_0)$, with the property that $\mu_0(H)=1$ and that the
observational noise $\{\eta_n\}_{n=1}^N$ is i.i.d.\ in
$\bbR^M$, independent of $u$, with $\eta_1$ distributed according to a
Gaussian measure $N(0,\gamma^2 I)$.

We let $C_0 = \pi^4 A^{-2}$ noting that if $u \sim \mu_0$, 
then $u \in H^s$ almost surely
for all $s<1$; in particular $u \in H$.  Thus $\mu_0(H)=1$ as
required.  The forcing in $f$ is taken to be $f=\nabla^{\perp}\Psi$,
where $\Psi=(2/\pi) ( \cos(\pi k_c \cdot x) + \sin(\pi k_s \cdot x)$ 
and $\nabla^{\perp}=J\nabla$ with $J$
the canonical skew-symmetric matrix, 
$k_c=(5,5)$ and $k_s=(-5,5)$.  Furthermore, we let the viscosity
$\nu=0.1$, and the interval between observations $h = \delta t =
0.05$, where $\delta t$ is the timestep of the numerical scheme, 
and $N=10$ so that the total time interval is $T=Nh=0.5$.  
In this regime, the nonlinearity is mild and the attractor is
a fixed point, similar to the linear case.  Nonetheless, the dynamics
are sufficiently nonlinear that the optimal proposal used in the 
previous section does not work.
We take $M=32^2$ evaluated at the gridpoints, such that the observations
include all numerically resolved, and hence observable, wavenumbers in
the system.  The observational noise standard deviation is taken to be
$\gamma = 3.2$.

\begin{figure}
 \includegraphics[width=0.5\textwidth]{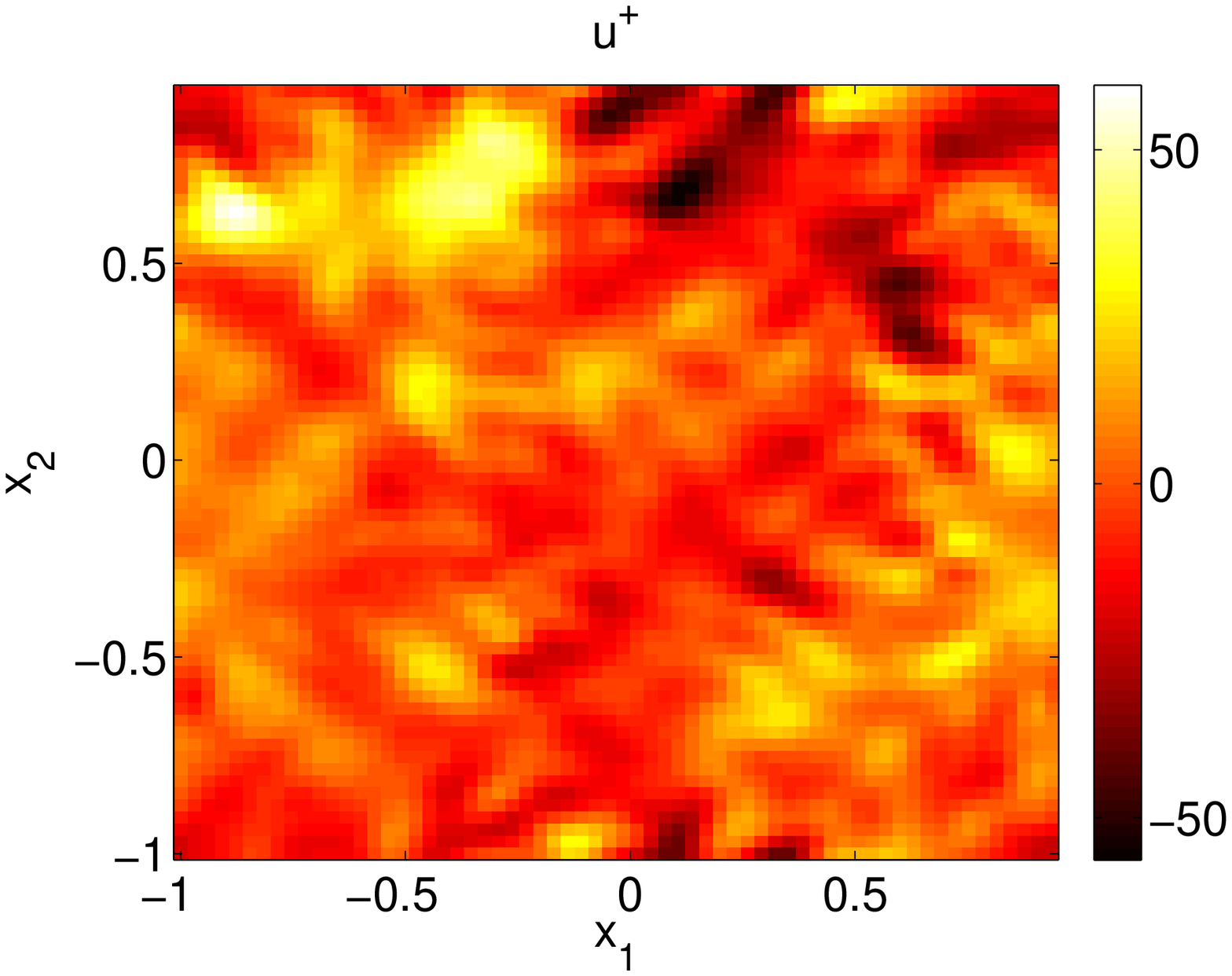}
 \includegraphics[width=0.5\textwidth]{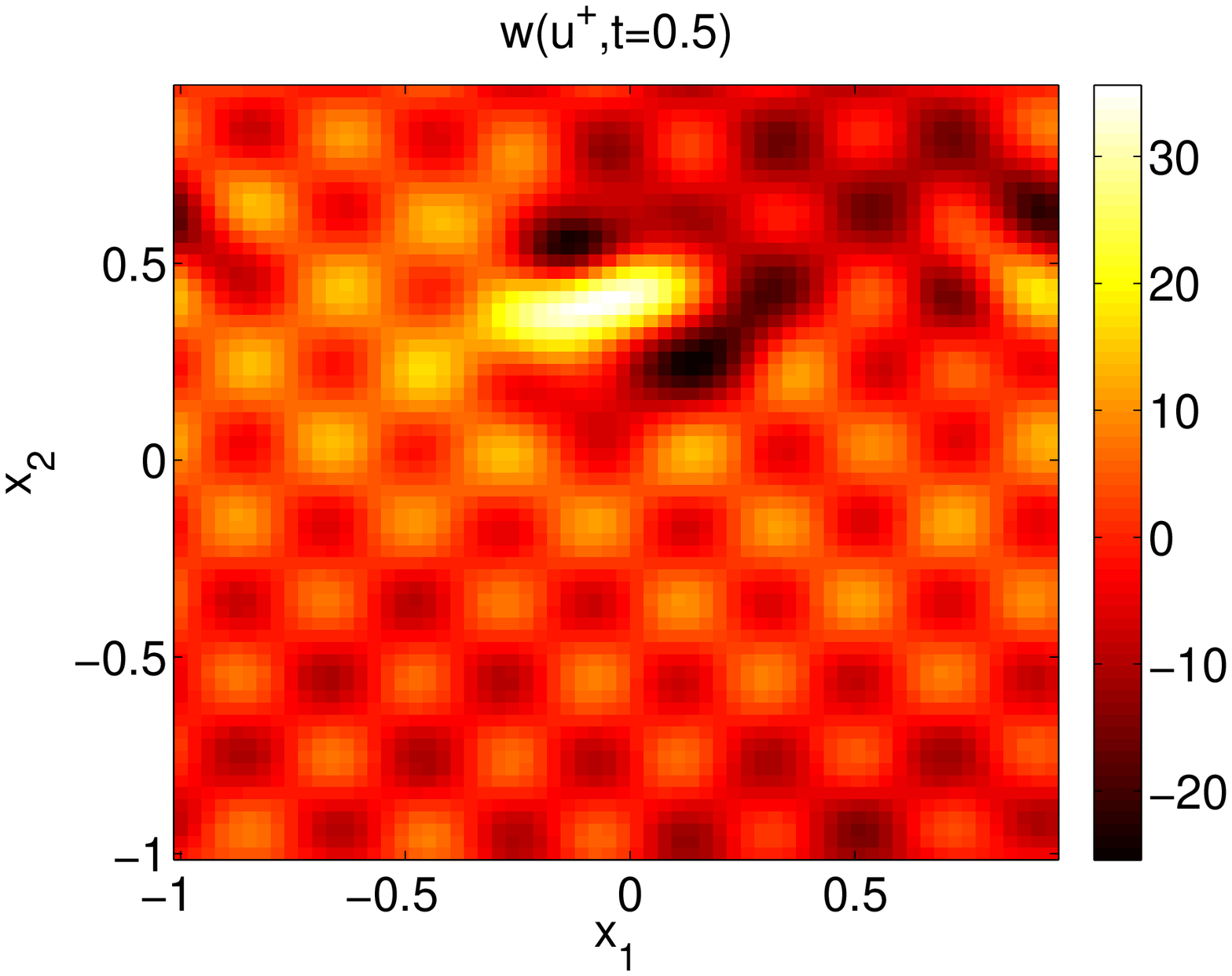}
 \includegraphics[width=0.5\textwidth]{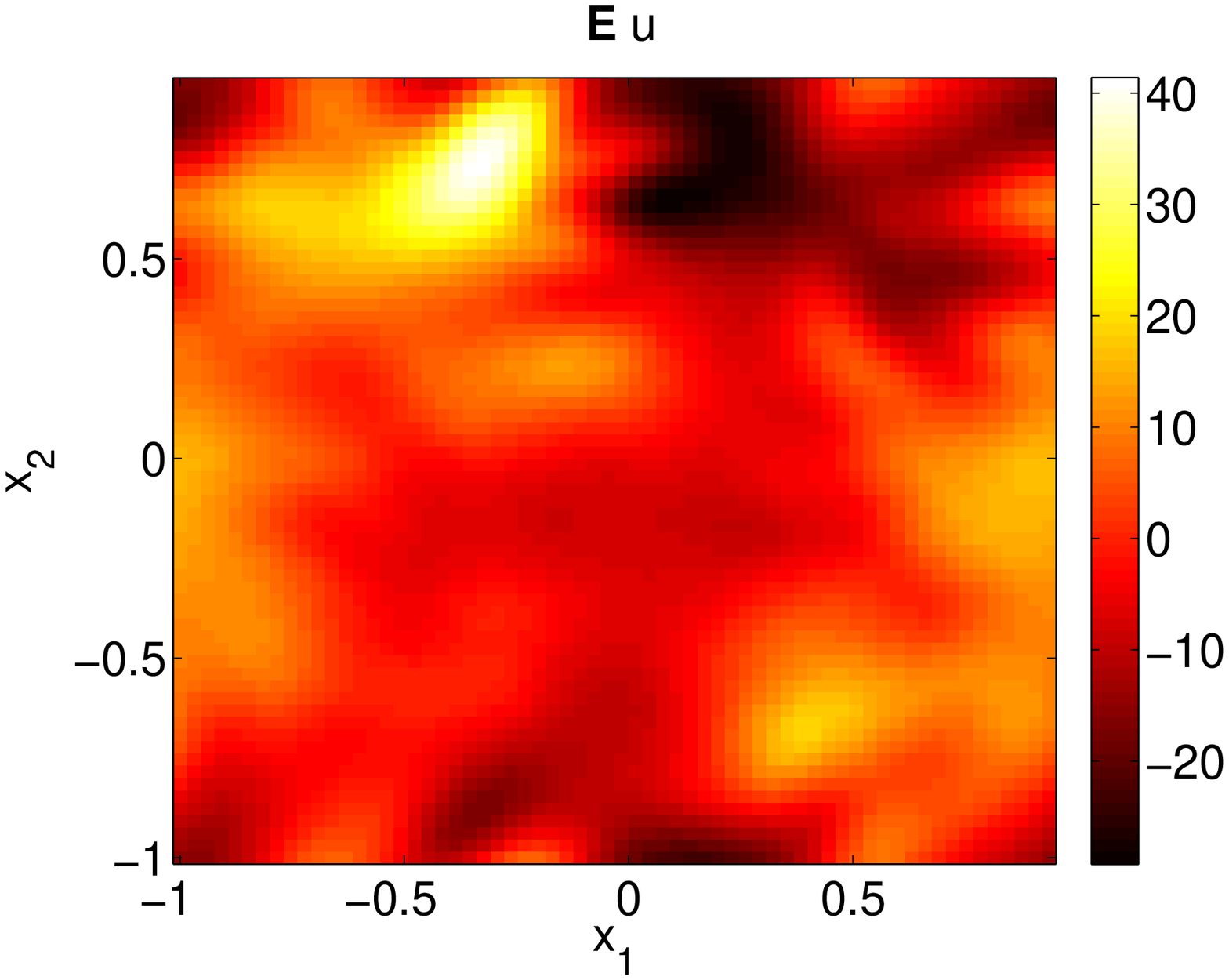}
 \includegraphics[width=0.5\textwidth]{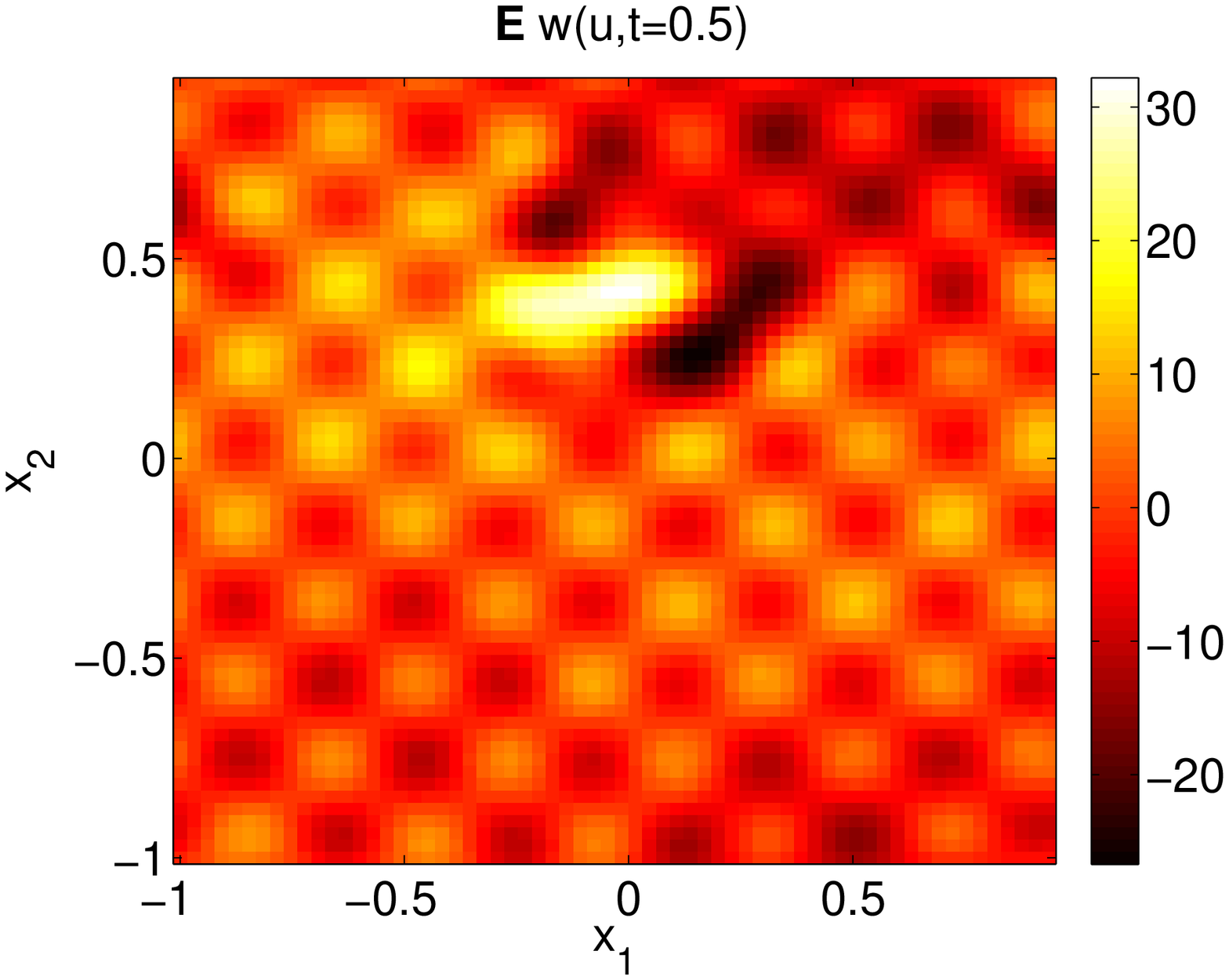}
\caption{Truth (top) and expected value of the posterior (bottom).
Initial condition is given to the left and solution at final time is
given to the right.}
\label{truth_mean}
\end{figure}

\begin{figure}
 \includegraphics[width=0.5\textwidth]{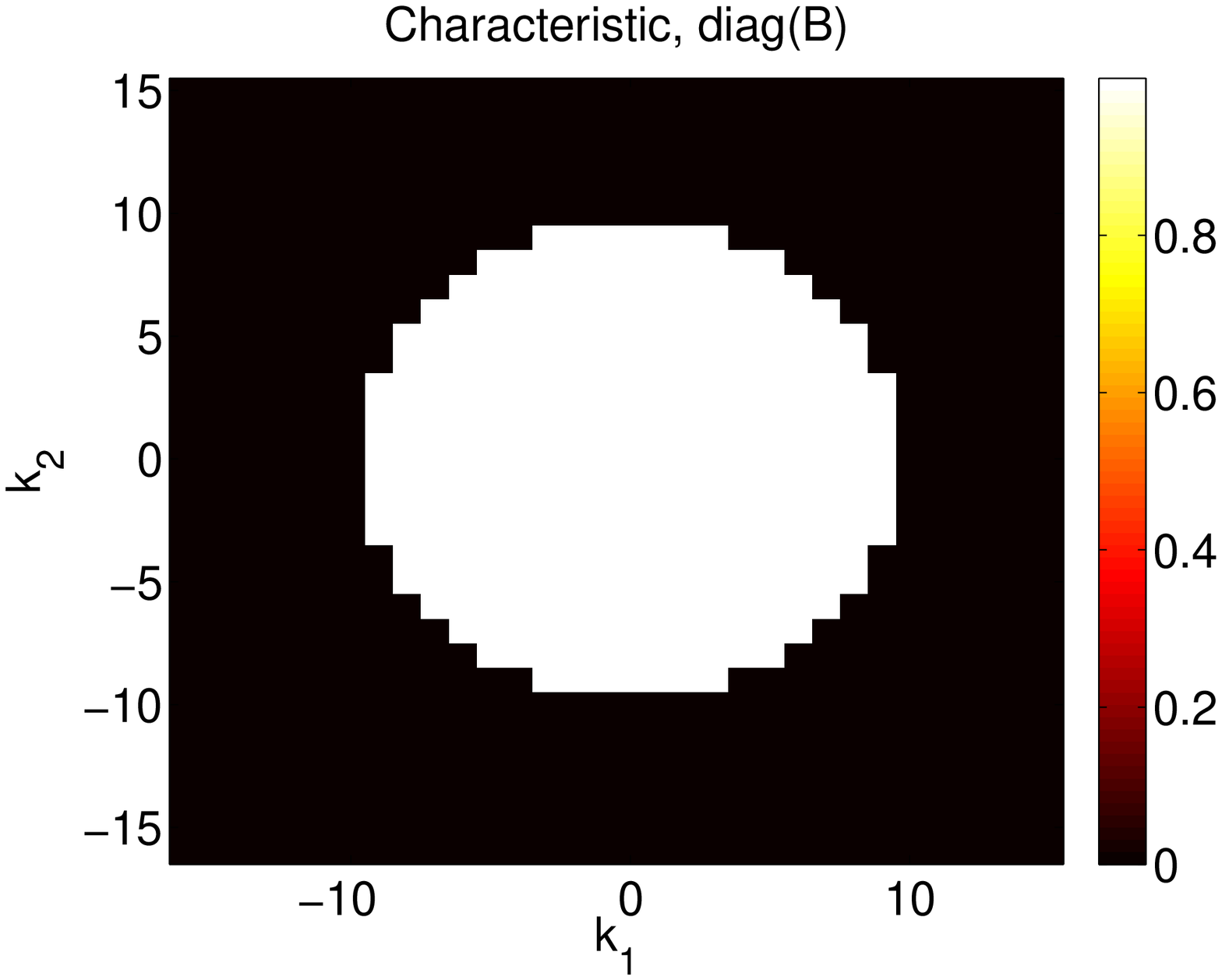}
 \includegraphics[width=0.5\textwidth]{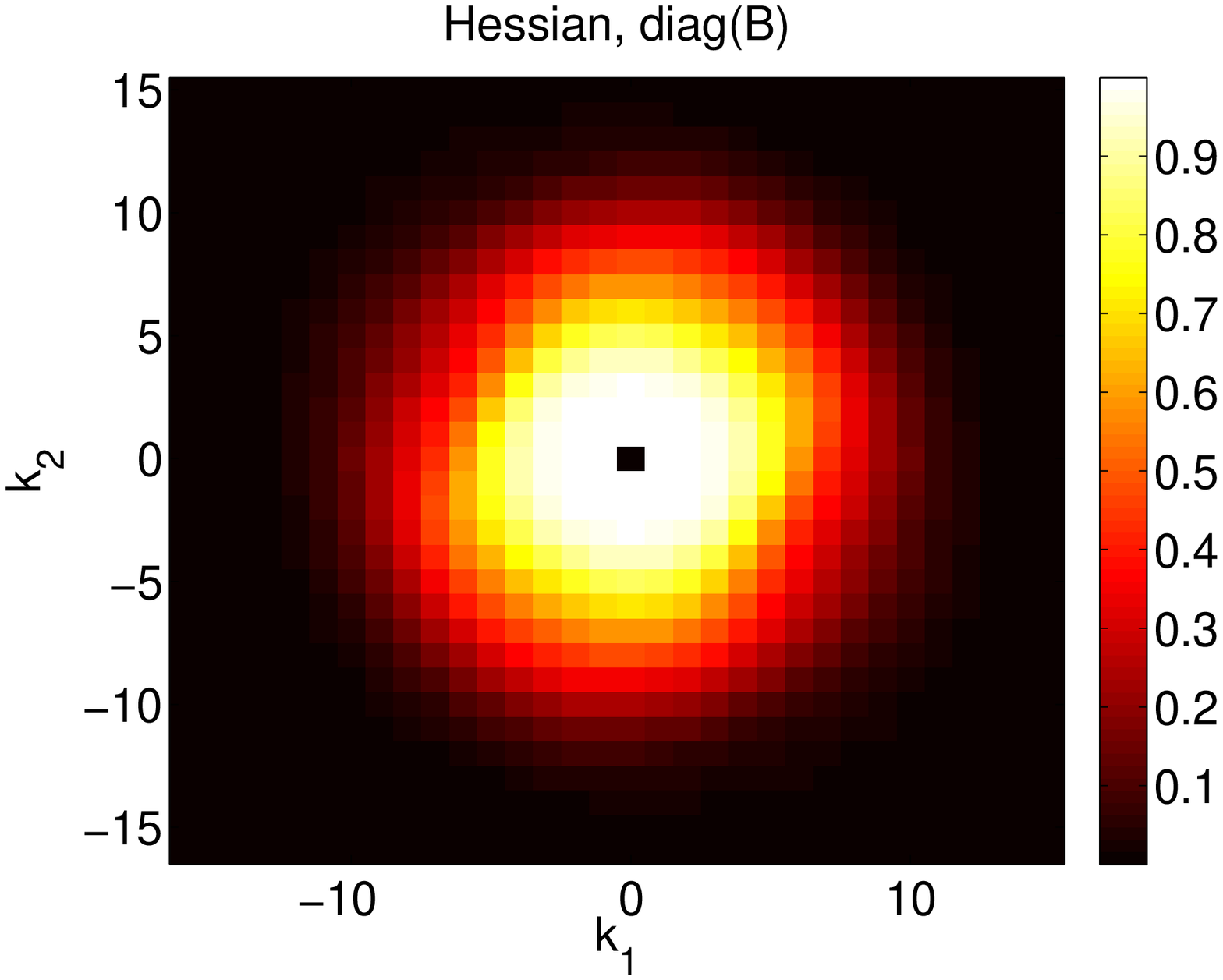}
\caption{The diagonal of the rescaling operator $B$ for C, 
given by \eqref{characteristic} and H, given by \eqref{optimalb}.}  
%One can consider O to be the case in which the 
%operator is given by the identity, i.e. all ones.}
\label{thebs}
\end{figure}

\begin{figure}
 \includegraphics[width=0.5\textwidth]{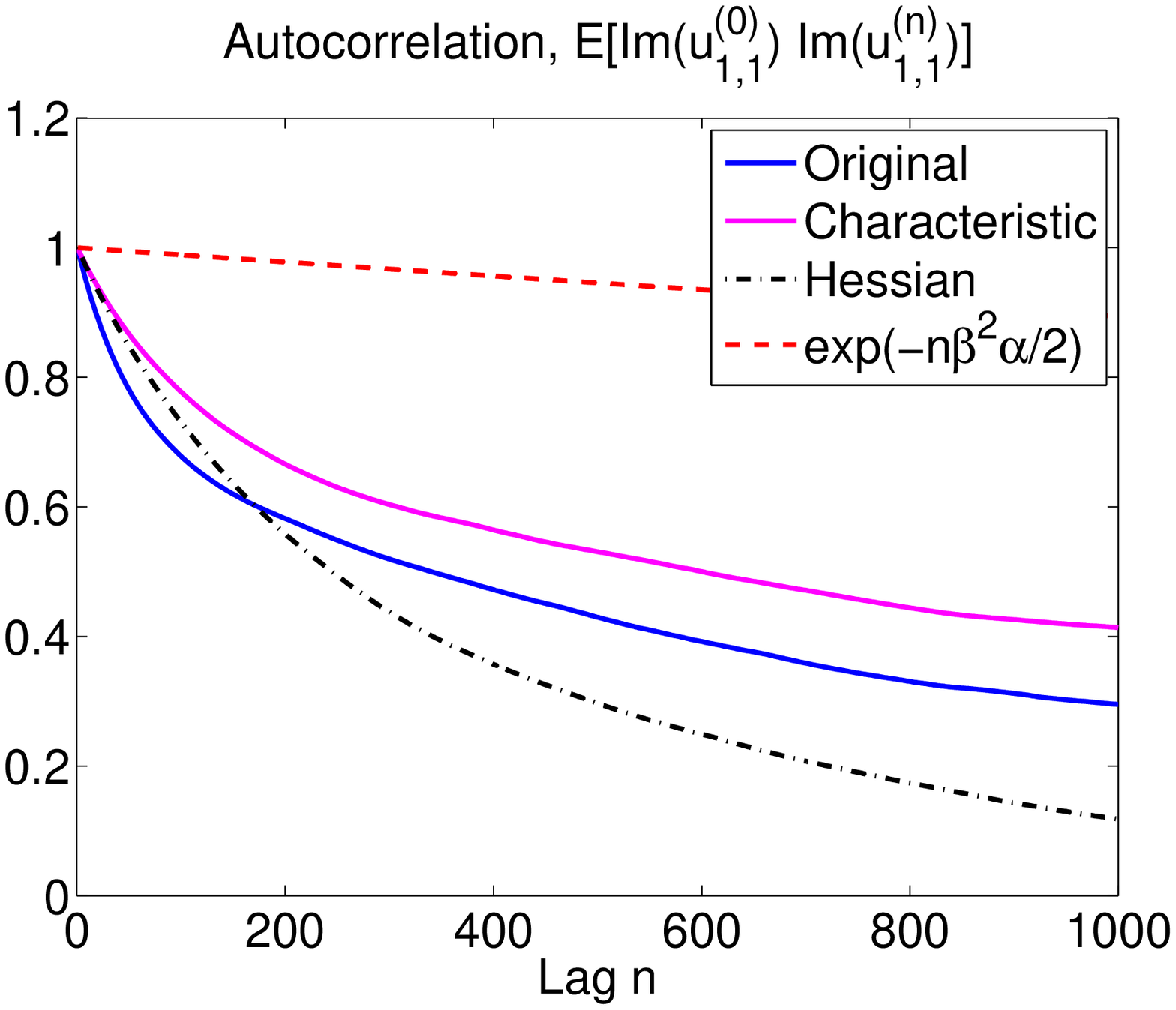}
 \includegraphics[width=0.5\textwidth]{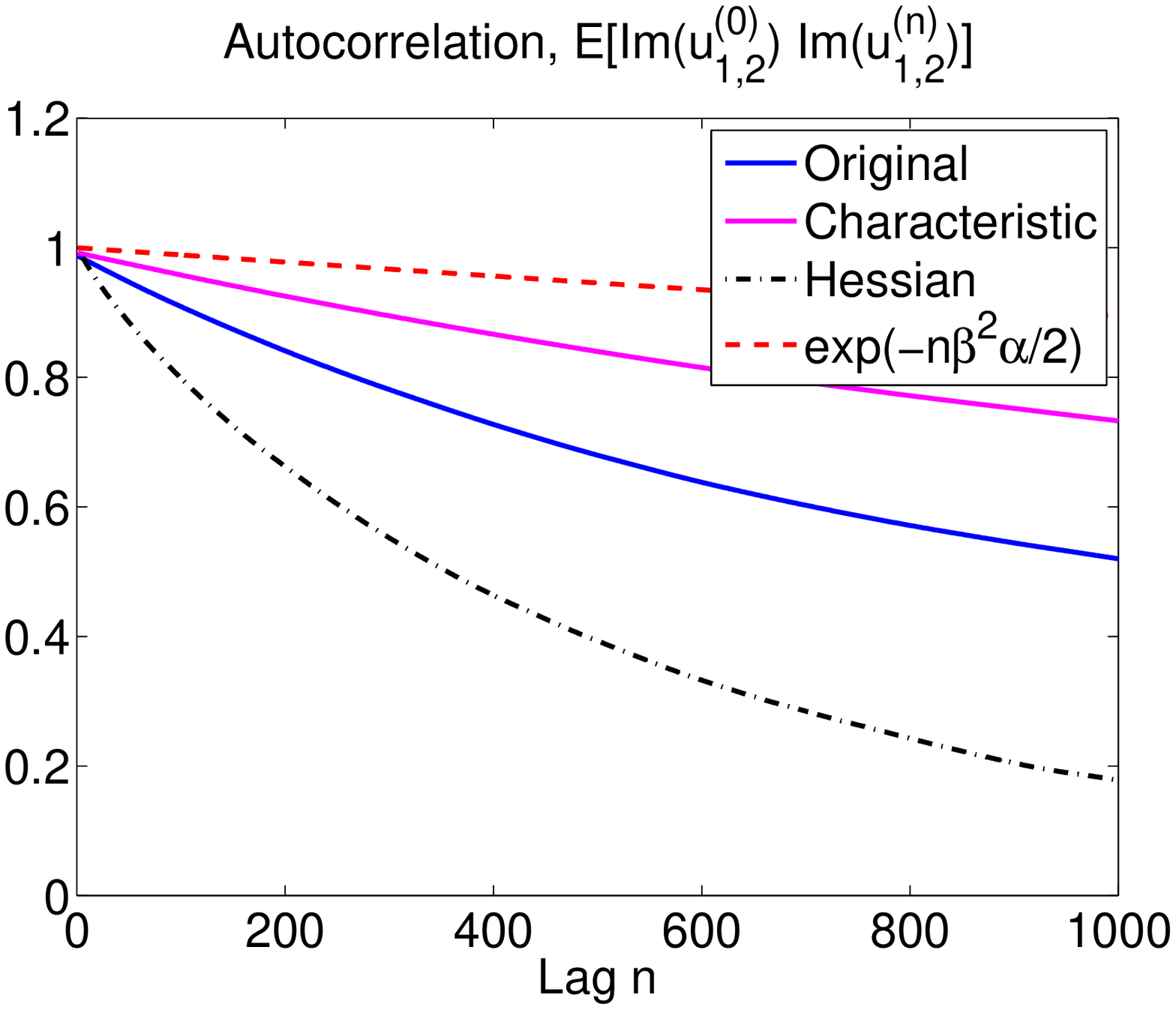}
 \includegraphics[width=0.5\textwidth]{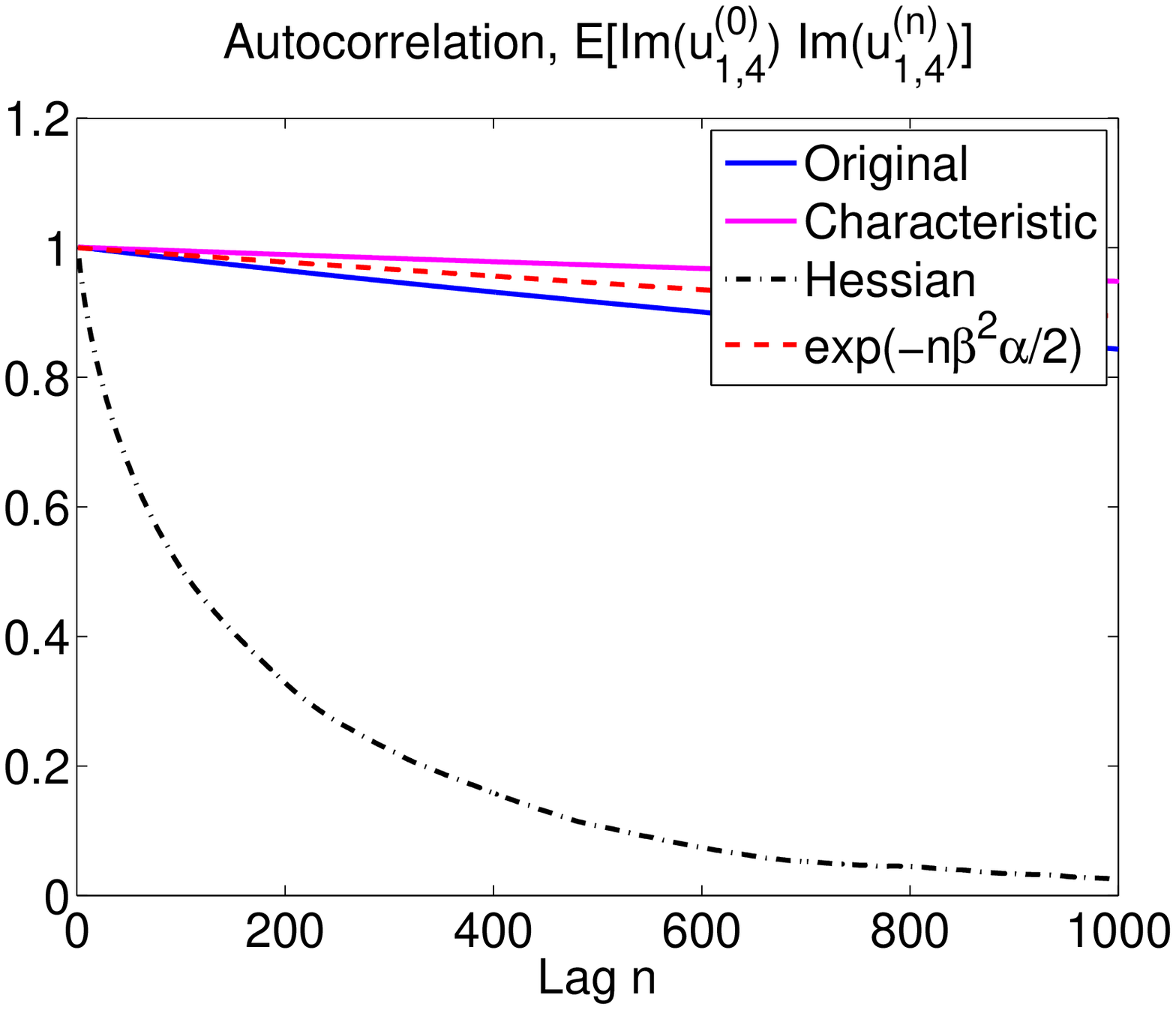}
 \includegraphics[width=0.5\textwidth]{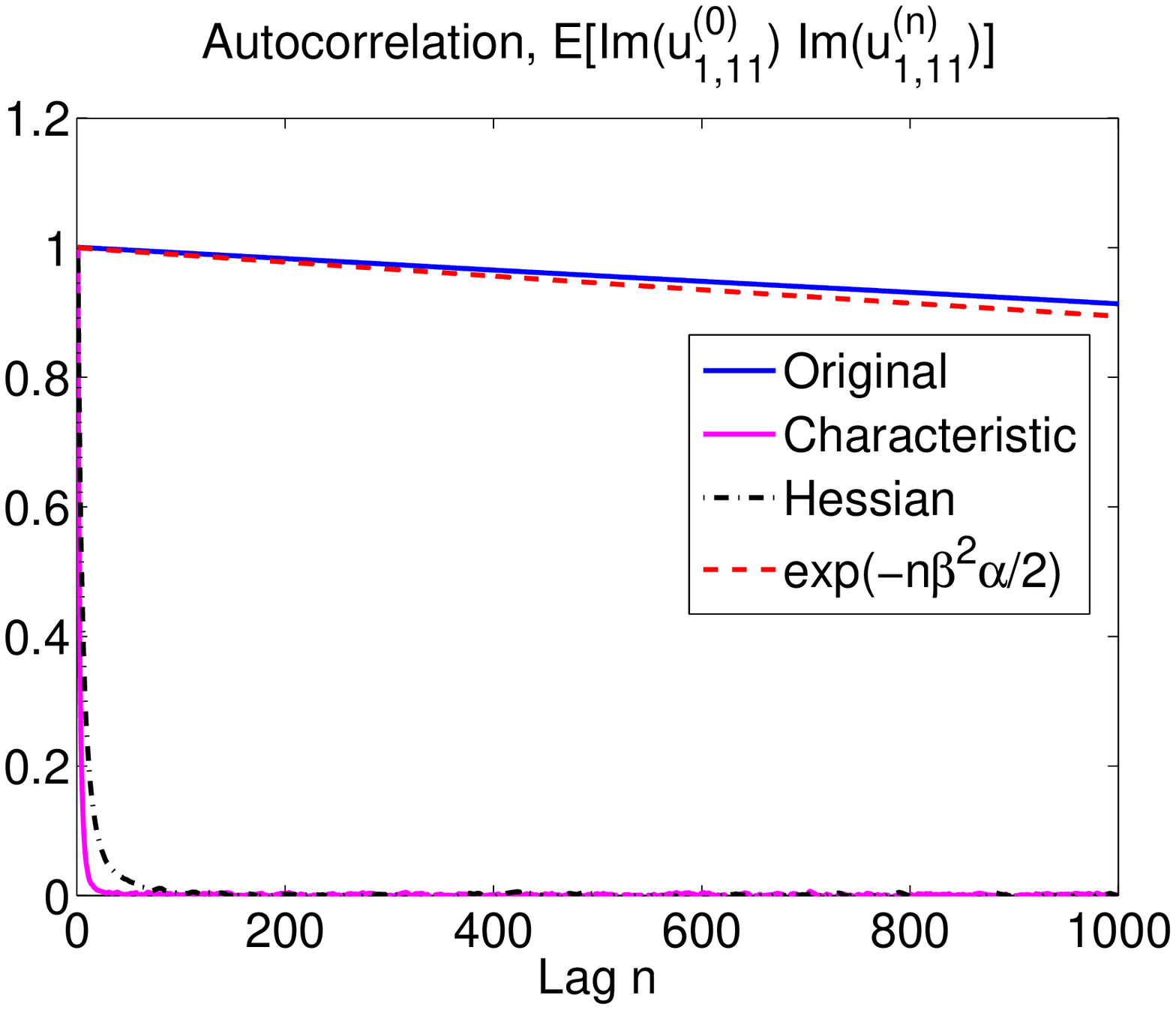}
\caption{The autocorrelation for each method O, C, and H,
for a variety of frequencies.  Also given for comparison 
is the exponential fit to the (acceptance lagged)
autocorrelation function for the proposal chain for O, 
i.e. $\exp (-n \beta^2 \bbE \alpha/2 )$.  This differs for each,
because $\beta$ and $\bbE \alpha$ differ.}
\label{autos}
\end{figure}

The truth and the mean vorticity initial conditions and solutions
at the end of the observation window are given in
Fig. \ref{truth_mean}. The truth (top panels)
clearly resembles a draw from a distribution with mean given
by the bottom panels.
Fig. \ref{thebs}  shows the diagonal of the rescaling operator $B$
(i.e. for $\beta=0$)
for C, given by (4.4) and H, given by (4.5). 
For O the operator is given by the identity, i.e. all one along the 
diagonal.
Some autocorrelation functions 
(approximated with one chain of $10^6$ each)
are shown in Fig. \ref{autos}.
 Also given for comparison 
is the exponential fit to the (acceptance lagged)
autocorrelation function for the proposal chain for O, 
i.e. $\exp (-n \beta^2 \bbE \alpha/2 )$.  This differs for each,
because $\beta$ and $\bbE \alpha$ differ.
In particular, for  O $\beta= 0.024$ and $\bbE\alpha\approx0.41$
(approximated by the sample mean), for
C $\beta= 0.015$ and $\bbE\alpha\approx0.32$, and  
for H $\beta=0.015$ and $\bbE\alpha\approx0.25$.
Presumably this accounts for the slight discrepancy in 
the autocorrelation of mode $u_{1,1}$, given in the top
left panel.  For the mode $u_{1,2}$ (top right), the difference 
becomes more pronounced between H and the other two.
In the bottom left it is clear that the mode $u_{1,4}$ 
decorrelates very quickly for H, while it decorrelates approximately
like the proposal chain for both other methods. The mode
$u_{1,11}$ (bottom right) is independently sampled for C, so has
decorrelation time $1/\bbE\alpha$.  It decorrelates almost as fast 
for H, and is decorrelating with the proposal chain for O.

\begin{figure}
 \includegraphics[width=0.5\textwidth]{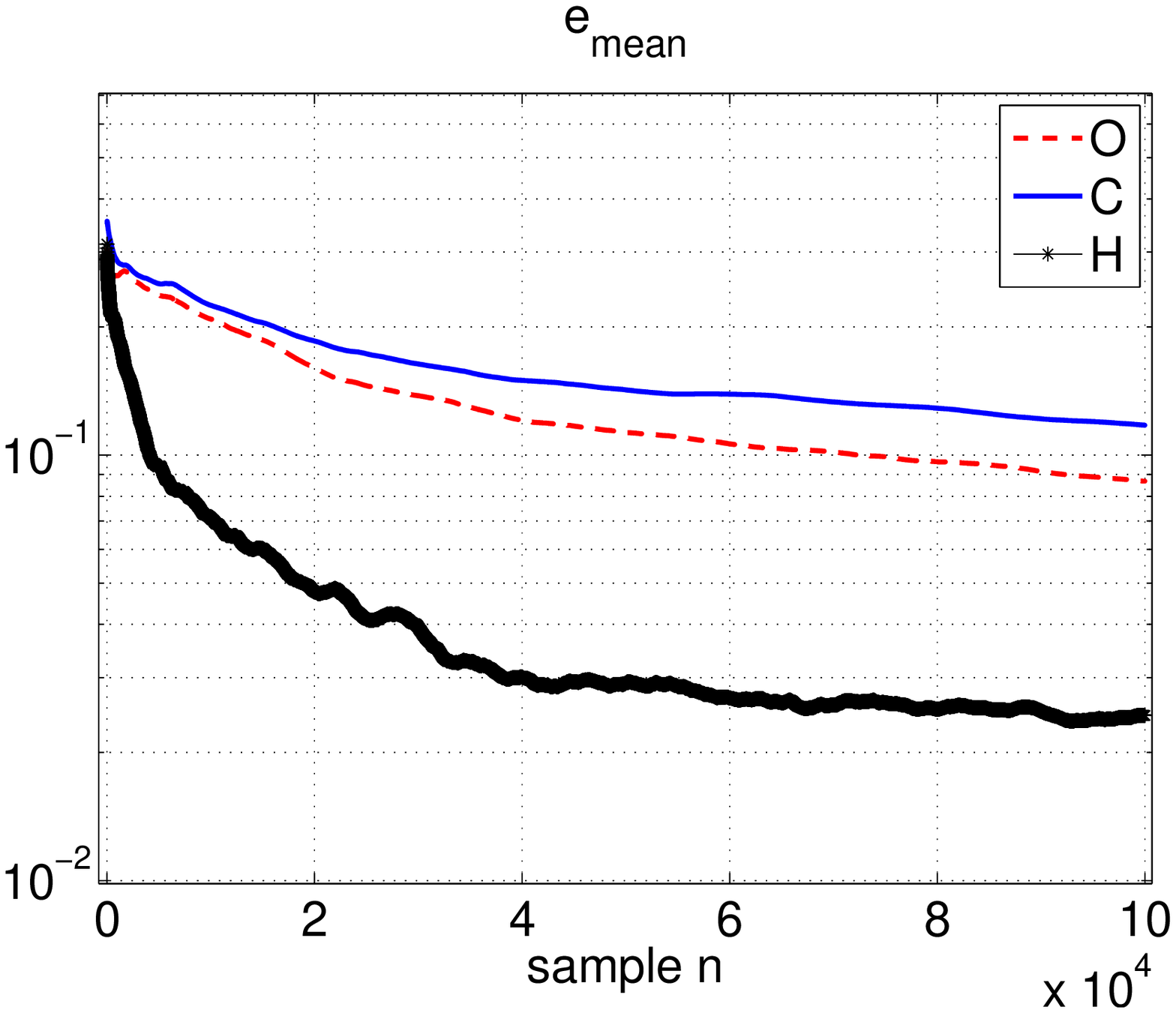}
 \includegraphics[width=0.5\textwidth]{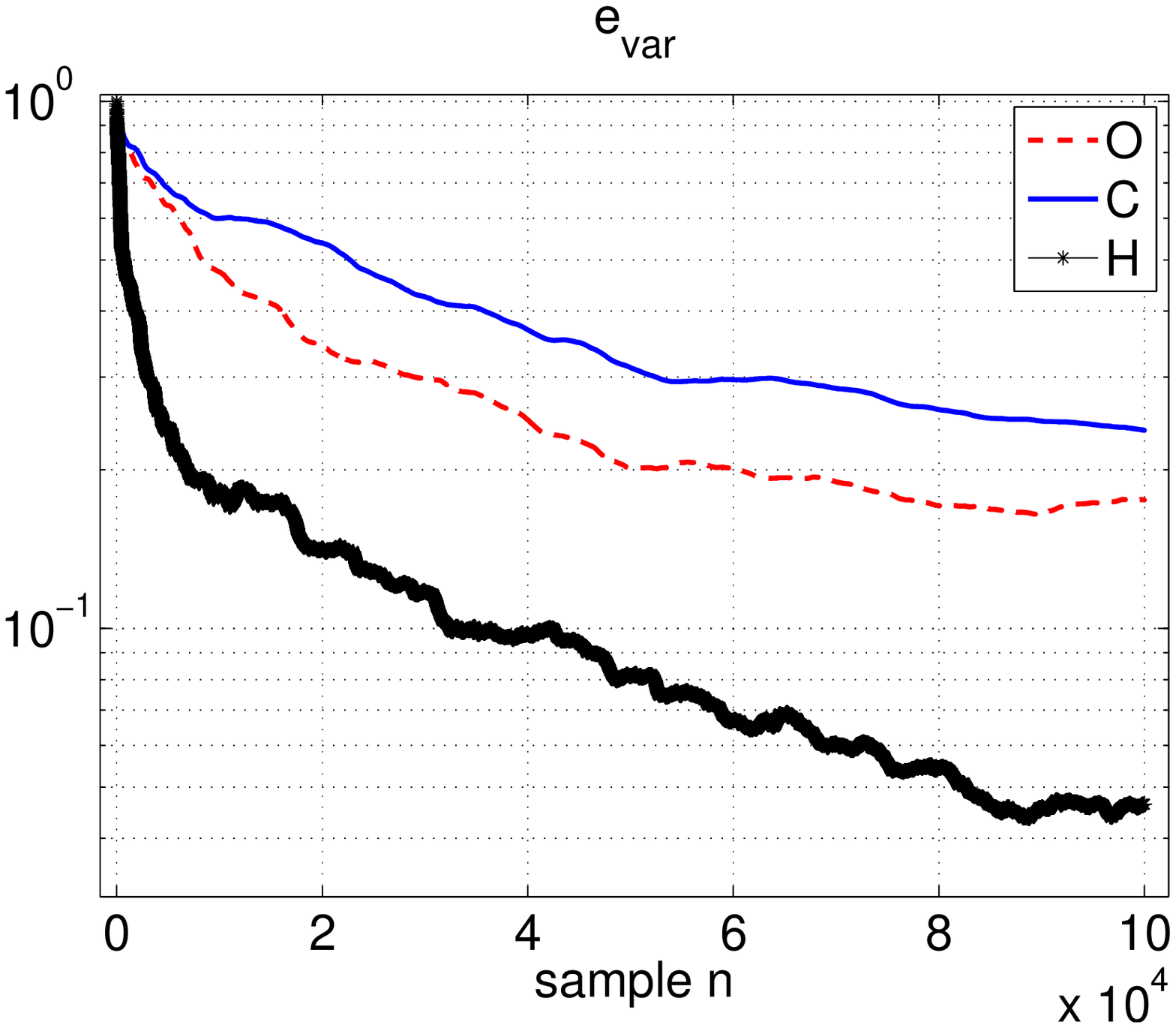}
\caption{The relative error in the mean (left) and variance (right)
with respect to the converged values as a function of sample number
for small sample sizes.}
\label{conv}
\end{figure}

\begin{figure}
 \includegraphics[width=0.3\textwidth]{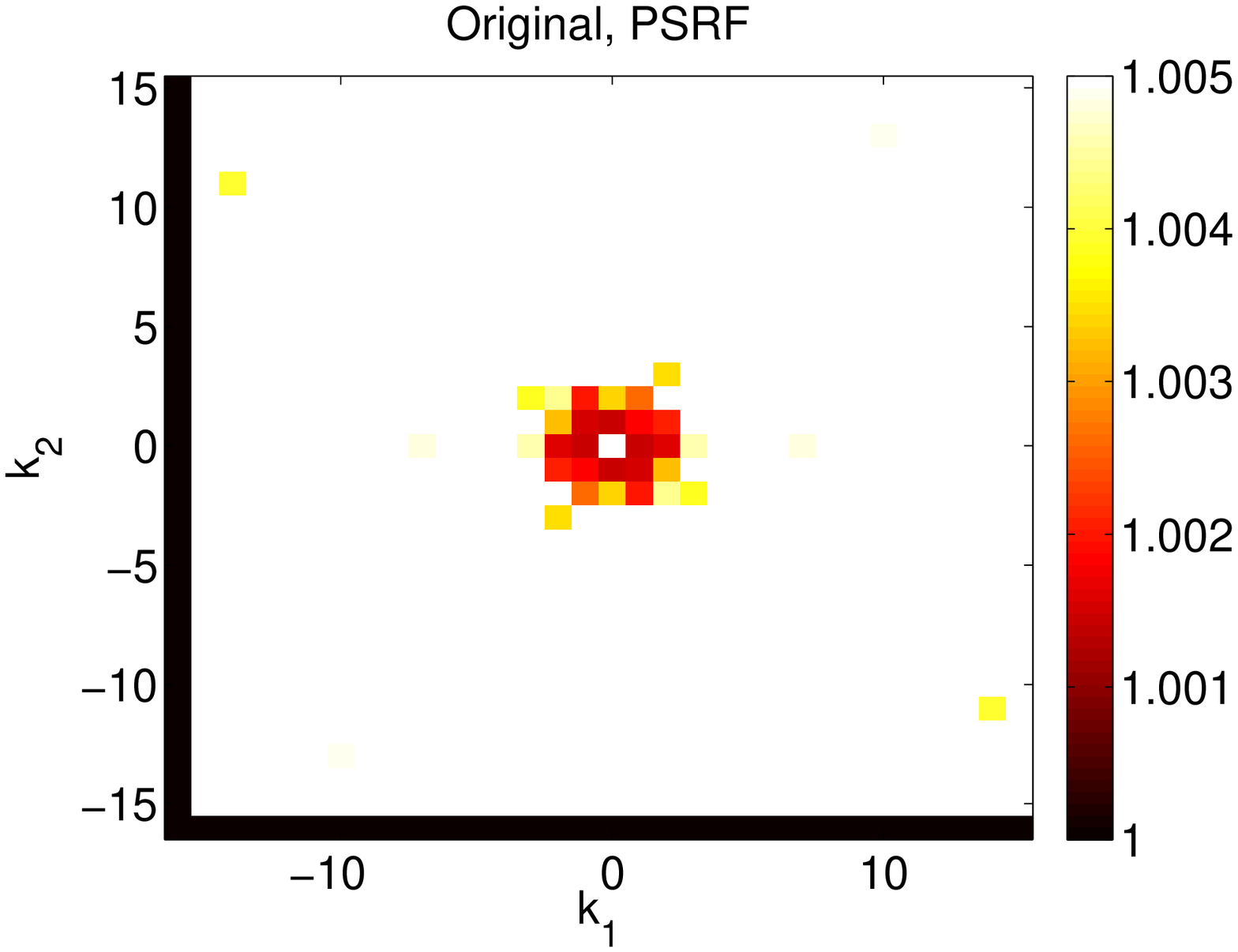}
 \includegraphics[width=0.3\textwidth]{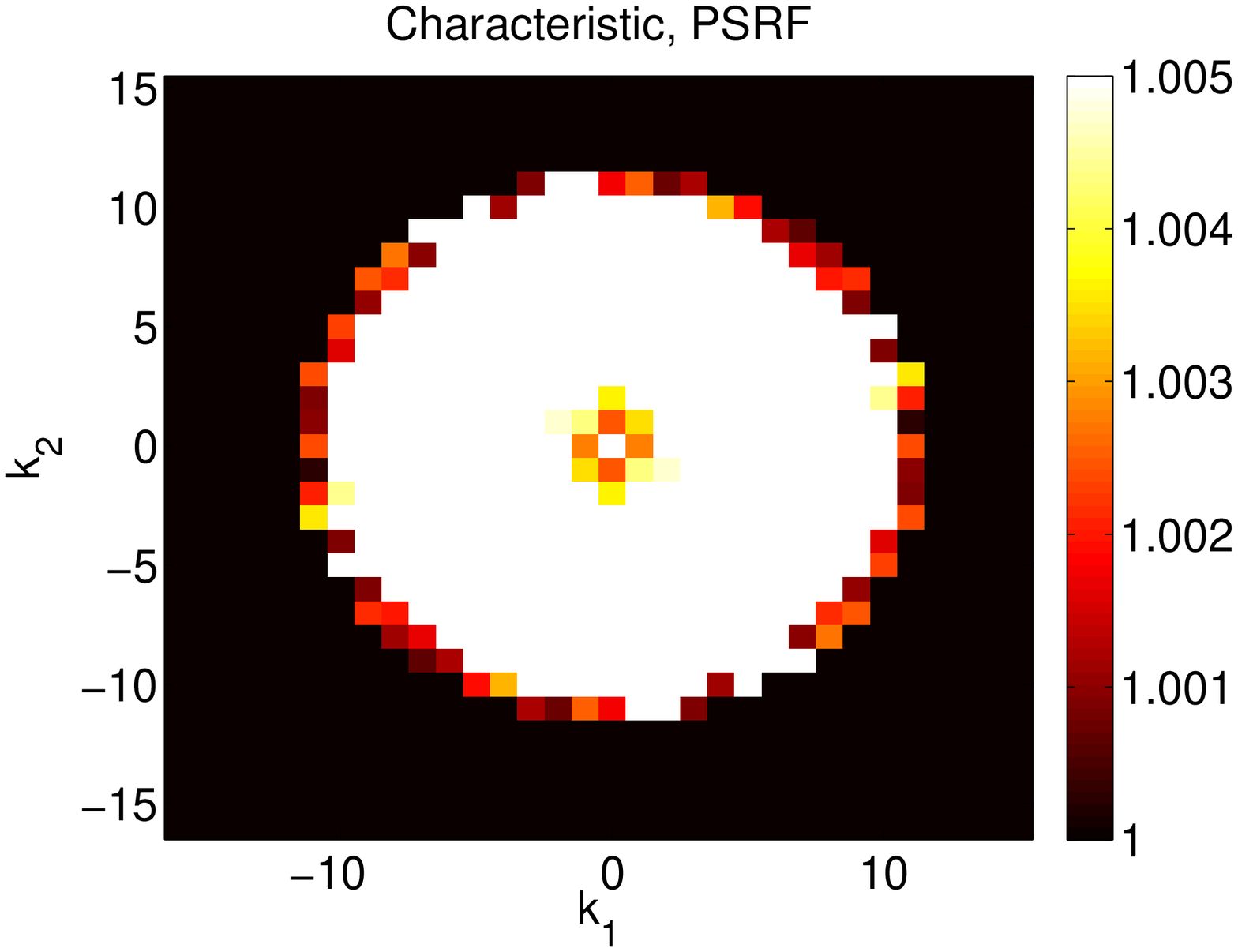}
 \includegraphics[width=0.3\textwidth]{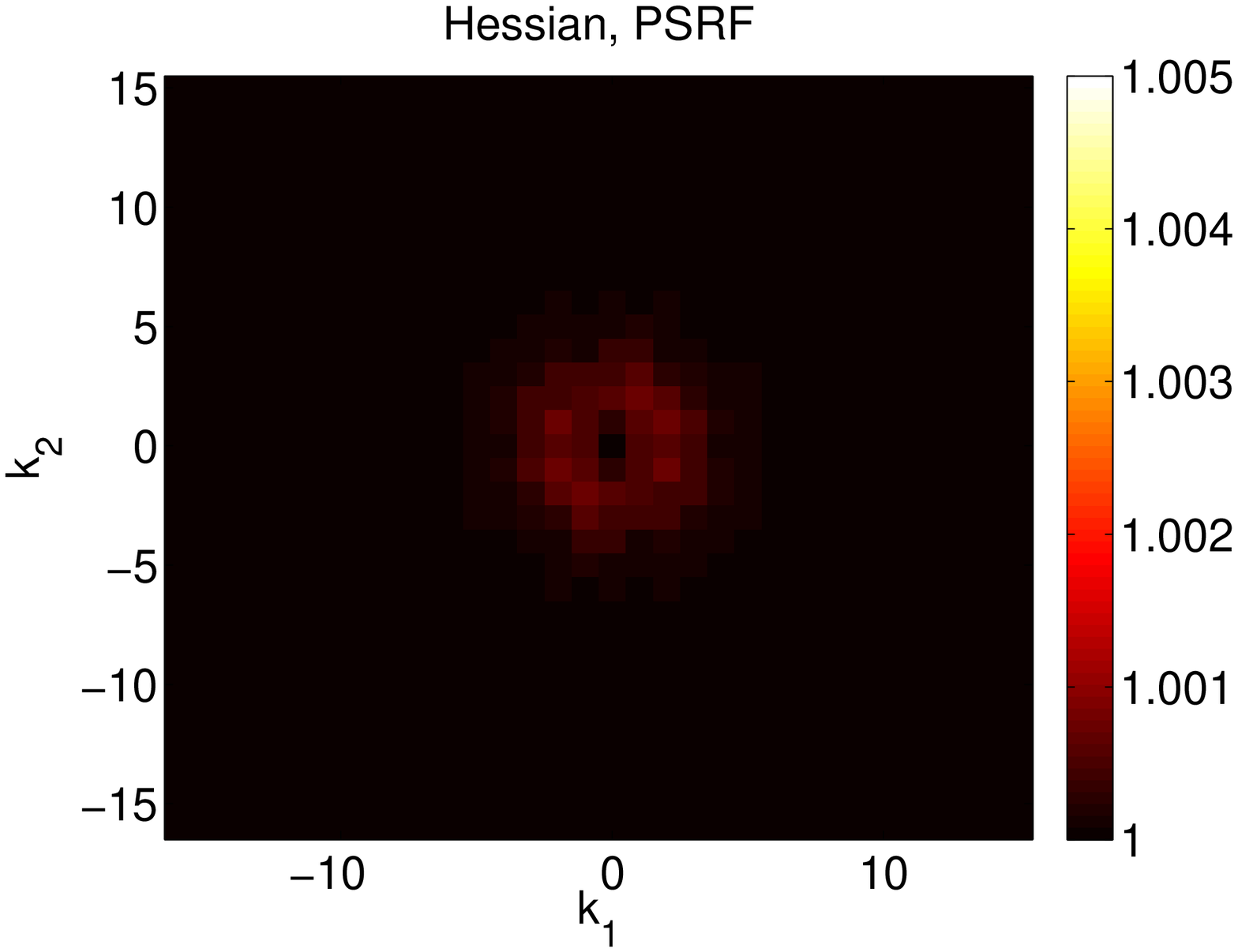}
\caption{The PSRF plots for O (left), C (middle) and H (right).  The
  images are saturated on the same scale so that H is visible. }
\label{psrf}
\end{figure}

Figure \ref{conv} shows the relative error in the mean (left) and 
variance (right) with 
respect to the converged values as a function of sample number 
for small sample sizes.  The accumulated benefit of H is very 
apparent here, yielding almost an order of magnitude improvement.
For each of the proposals, we run $16$ chains of length $10^6$, 
%so that (i) all methods produce a very good collection of samples, and
%(ii) we can 
both for the benchmark mean and variance used in Fig. \ref{conv}, and to
compare the so-called potential scale reduction
factor (PSRF) convergence diagnostic \cite{brooks1998general}. 
In Fig. \ref{psrf} we plot the PSRF for each method on the same color
scale from $[1,1.005]$.
Although it cannot be seen for O(left) and C(middle) since the values
of most modes are saturated on this scale, 
every mode for every method is clearly converged according to
the PSRF, which should be smaller than 1.1 and eventually converges to 1. 
We can see for O that the low frequencies, where the mean and 
uncertainty are concentrated, converge 
faster (as indicated in \ref{autos}), and other modes converge 
more or less uniformly slowly.
C clearly outperforms O in the high frequencies outside the
truncation radius (where the mean
and uncertainty are negligible) and is comparable for the 
smallest frequencies, but may even be worse than O
for the intermediate frequencies.  This may be explained by the 
different $\beta$ and $\bbE\alpha$.
Again, H quite clearly outperforms both of the other two methods.  
Note that these results are for particular parameter choices, 
and methods of choosing the parameters $\beta$, $k_c$,
$\zeta$, and may be sensitive to their choice.

%$e^c_m = 0.0121$, $e^o_m = 0.0073$, $e^c_{m2}=0.0025$, $e^o_%{m2}=0.0011$,
%$e^c_{v}=0.003$, $e^o_{v}=0.0013$

\section{Conclusions}
\label{conclusion}

This manuscript introduces and investigates a new class of proposals
for function-space MCMC, giving rise to a new class of algorithms.
These proposals involve an operator which 
appropriately rescales the proposal step-sizes in different directions 
in order to decrease autocorrelation and hence improve convergence 
rate.  
For the curvature-inspired rescaling H introduced here, there are two design
parameters which need to be hand-chosen.  But, it seems to be effective
to choose $\zeta$ sufficiently small and then allow $\beta$ to be
chosen during a transient adaptive phase, as in the other methods.
It should be possible to 
achieve the same result with one design parameter.
%However, there are many 
%for periodic updates or distributions which are closer
%to Gaussian  may work
%and is the topic of ongoing research.
Furthermore, 
there exists a whole range of more sophisticated models 
between those investigated 
here and the more sophisticated but expensive 
RMHMC \cite{girolami2011riemann} and 
stochastic Newton \cite{martin2012stochastic}.  This gives rise to 
exciting opportunities for future investigation.  It will be of
great interest to establish rigorous theoretical results, such as 
geometric ergodicity, for these methods.
\\
\\
\noindent{\bf Acknowledgements}
{The author would like to thank 
EPSRC and ONR for financial support.
I also thank The Mathematics Institute and Centre for Scientific
Computing at Warwick University, as well as 
The Institute for Computational
and Experimental Research in Mathematics and 
The Center for Computation and Visualization at
Brown University for supplying valuable computation
time.  I would like to extend thanks to 
Tiangang Cui, Youssef Marzouk, and Gareth Roberts
for interesting discussions related to these results, and
also to the referee for a careful read and 
insightful suggestions for improvement. 
Finally, I thank Andrew Stuart for interesting discussions
related to these results and for valuable input in the revision 
of this manuscript.}

% \bibliographystyle{elsarticle-num}
% %\bibliographystyle{plain}
% \bibliography{mybib}

\begin{thebibliography}{10}
\expandafter\ifx\csname url\endcsname\relax
  \def\url#1{\texttt{#1}}\fi
\expandafter\ifx\csname urlprefix\endcsname\relax\def\urlprefix{URL }\fi
\expandafter\ifx\csname href\endcsname\relax
  \def\href#1#2{#2} \def\path#1{#1}\fi

\bibitem{Stuart10}
A.~Stuart, Inverse problems: a {B}ayesian approach, Acta Numer. 19 (2010)
  451--559.

\bibitem{CRSW12}
S.~Cotter, G.~Roberts, A.~Stuart, D.~White, {MCMC} methods for functions:
  modifying old algorithms to make them faster, Arxiv preprint arXiv:1202.0709.

\bibitem{girolami2011riemann}
M.~Girolami, B.~Calderhead, Riemann manifold {L}angevin and {H}amiltonian
  {M}onte {C}arlo methods, Journal of the Royal Statistical Society: Series B
  (Statistical Methodology) 73~(2) (2011) 123--214.

\bibitem{martin2012stochastic}
J.~Martin, L.~Wilcox, C.~Burstedde, O.~Ghattas, A stochastic newton mcmc method
  for large-scale statistical inverse problems with application to seismic
  inversion, SIAM Journal on Scientific Computing 34~(3) (2012) 1460--1487.

\bibitem{BRSV08}
A.~Beskos, G.~O. Roberts, A.~M. Stuart, J.~Voss, {MCMC} methods for diffusion
  bridges, Stochastic Dynamics 8~(3) (2008) 319--350.
\newblock \href {http://dx.doi.org/10.1142/S0219493708002378}
  {\path{doi:10.1142/S0219493708002378}}.

\bibitem{GGR97}
G.~Roberts, A.~Gelman, W.~Gilks, Weak convergence and optimal scaling of random
  walk {M}etropolis algorithms, Ann. Appl. Prob. 7 (1997) 110--120.

\bibitem{robtwe96}
G.~Roberts, R.~Tweedie, Exponential convergence of {L}angevin distributions and
  their discrete approximations, Bernoulli 2~(4) (1996) 341--363.

\bibitem{MRRT53}
N.~Metropolis, R.~Rosenbluth, M.~Teller, E.~Teller, Equations of state
  calculations by fast computing machines, J. Chem. Phys. 21 (1953) 1087--1092.

\bibitem{Has70}
W.~Hastings, Monte carlo sampling methods using markov chains and their
  applications, Biometrika 57 (1970) 97--109.

\bibitem{Tie}
L.~Tierney, A note on {M}etropolis-{H}astings kernels for general state spaces,
  Ann. Appl. Probab. 8~(1) (1998) 1--9.

\bibitem{HSV12}
M.~Hairer, A.~Stuart, S.~Vollmer, Spectral gaps for a {M}etropolis-{H}astings
  algorithm in infinite dimensions, Arxiv preprint arXiv:1112.1392.

\bibitem{roberts2007coupling}
G.~Roberts, J.~Rosenthal, Coupling and ergodicity of adaptive markov chain
  monte carlo algorithms, Journal of applied probability 44~(2) (2007)
  458--475.

\bibitem{book:Robinson2001}
J.~C. Robinson, Infinite-{D}imensional {D}ynamical {S}ystems, Cambridge Texts
  in Applied Mathematics, Cambridge University Press, Cambridge, 2001.

\bibitem{brooks1998general}
S.~Brooks, A.~Gelman, General methods for monitoring convergence of iterative
  simulations, Journal of Computational and Graphical Statistics 7~(4) (1998)
  434--455.

\bibitem{ourmap}
M.~Dashti, K.J.H.~Law, A.M.~Stuart and J.~Voss, 
MAP estimators and posterior consistency in Bayesian nonparametric inverse
problems. Accepted for publication in Inverse Problems (2013).  http://arxiv.org/abs/1303.4795 .

\bibitem{TC}
T.~Cui, K.J.H.~Law, and Y.~Marzouk, Likelihood-informed sampling of Bayesian nonparametric inverse problems.  In preparation (2013).

\bibitem{bckj}
A.~Beskos, A.~Jasra, N.~Kantas, (Submitted, 2013). http://arxiv.org/abs/1307.6127 .


\end{thebibliography}

\end{document}